\documentclass[12pt,usenames,dvipsnames]{article}

\usepackage{latexsym}
\usepackage{amssymb,amsfonts,amsmath}
\usepackage{graphicx} 
\usepackage{indentfirst}
\usepackage{bbm}
\usepackage{amssymb}
\usepackage{verbatim}
\usepackage{amsmath, amsthm,amssymb}
\usepackage{mathrsfs}
\usepackage{hyperref}
\usepackage{amsfonts}
\usepackage{dsfont}
\usepackage{cite}
\usepackage{xcolor}
\usepackage{enumerate}
\usepackage{cleveref}

\parskip=\medskipamount

\newcommand {\cD}{{\cal D}}
\newcommand {\cE}{{\cal E}}

\newcommand {\cK}{{\cal K}}

\newcommand {\cM}{{\cal M}}
\newcommand {\cN}{{\cal N}}

\newcommand {\cR}{{\cal R}}

\newcommand {\cT}{{\cal T}}

\newcommand {\cW}{{\cal W}}
\newcommand {\cX}{{\cal X}}

\def\a{\alpha}

\def\b{\beta}
\def\c{\chi}
\def\d{\delta}

\def\g{\gamma}

\def\k{\kappa}

\def\m{\mu}

\def\q{\theta}
\def\r{\rho}
\def\s{\sigma}

\def\D{\Delta}
\def\F{\Phi}
\def\J{\Psi}
\def\L{\Lambda}
\def\O{\Omega}

\def\S{\Sigma}
\def\U{\Upsilon}

\def\rd{{\rm d}}
\def\ri{{\rm i}}
\def\re{{\rm e}}

\newcommand{\ad}{{\dot{\alpha}}}                           %
\newcommand{\bd}{{\dot{\beta}}}                            %
\newcommand{\ve}{\varepsilon}                            %
\newcommand{\cDB}{{\bar\cD}}                            %

\renewcommand{\aa}{{\a\ad}}

\newcommand{\pa}{\partial}                           %
\newcommand{\hf}{\frac12}

\newcommand{\vf}{\varphi}

\newcommand{\be}{\begin{equation}}
\newcommand{\ee}{\end{equation}}
\newcommand{\bea}{\begin{eqnarray}}
\newcommand{\eea}{\end{eqnarray}}
\newcommand{\non}{\nonumber}
\newcommand{\1}{{\underline{1}}}
\newcommand{\2}{{\underline{2}}}

\newcommand{\bm}[1]{\mbox{\boldmath$#1$}}

\def\double #1{#1{\hbox{\kern-2pt $#1$}}}

\newcommand{\gd}{{\dot\g}}
\newcommand{\dd}{{\dot\d}}

\newcommand{\ts}{{\tilde{\s}}}

\newcommand{\sba}{{\bar{\s}}}

\newcommand{\teb}{{\bar{\theta}}}

\newcommand{\Nabla}{\bm{\nabla}}
\newcommand{\bNabla}{\bar{\bm{\nabla}}}

\newif\ifdtup

\newcommand{\bsubeq}{\begin{subequations}}
\newcommand{\esubeq}{\end{subequations}}

\numberwithin{equation}{section}

\newcommand{\sSU}{\mathsf{SU}}
\newcommand{\sSL}{\mathsf{SL}}

\newcommand{\sSO}{\mathsf{SO}}

\newcommand{\sU}{\mathsf{U}}

\begin{document}

\begin{titlepage}
\begin{flushright}
January, 2025 \\
Revised version: April, 2025
\end{flushright}
\vspace{5mm}

\begin{center}
{\Large \bf 
${\mathcal N}=3$ nonlinear multiplet and supergravity
}
\end{center}

\begin{center}

{\bf Sergei M. Kuzenko and Emmanouil S. N. Raptakis} \\
\vspace{5mm}

\footnotesize{
{\it Department of Physics M013, The University of Western Australia\\
35 Stirling Highway, Perth W.A. 6009, Australia}}  
~\\
\vspace{2mm}
~\\
Email: \texttt{ 
sergei.kuzenko@uwa.edu.au, emmanouil.raptakis@uwa.edu.au}\\
\vspace{2mm}

\end{center}

\begin{abstract}
\baselineskip=14pt

We propose an ${\mathcal N}=3$ nonlinear multiplet coupled to conformal supergravity and use it to formulate the equations of motion for ${\mathcal N} = 3$ Poincar\'e supergravity. These equations, which are naturally described in a new curved supergeometry with structure group $\mathsf{SL}(2,\mathbb{C})$, imply that the ${\mathcal N} = 3$ super-Bach tensor vanishes, and thus every solution of Poincar\'e supergravity is a solution of conformal supergravity. The aforementioned superspace formulation, which we refer to as $\mathcal N=3$ Einstein superspace, is described in terms of two dimension-$1/2$ superfields: (i) the super-Weyl spinor $W_\alpha$; and (ii) a spinor isospinor $\chi_\alpha^i$. 
\end{abstract}
\vspace{5mm}

\vfill

\vfill
\end{titlepage}

\newpage
\renewcommand{\thefootnote}{\arabic{footnote}}
\setcounter{footnote}{0}

\tableofcontents{}
\vspace{1cm}
\bigskip\hrule

\allowdisplaybreaks

\section{Introduction}

Pure $\cN=2$ supergravity in four dimensions was constructed by Ferrara and van Nieuwenhuizen in 1976 \cite{FvN}. It may be viewed to be a realisation of Einstein's dream of unifying gravity and electromagnetism. 
In 1979, the first off-shell formulation for this theory was proposed by Fradkin and Vasiliev \cite{Fradkin:1979cw} and soon after by de Wit, van Holten and Van Proeyen \cite{deWit:1979dzm}.\footnote{These works extended the earlier linearised results of \cite{Fradkin:1979as, deWit:1979xpv}.}
The latter authors also provided a reformulation of the theory \cite{deWvanHvanP} within the framework of the $\cN=2$ superconformal multiplet calculus developed in \cite{deWit:1979dzm, BdeRdeW, deWvanHvanP, deWPV}.
 Its specific feature is the use of the so-called nonlinear multiplet  \cite{deWvanHvanP} as one of the superconformal compensators.
 We recall that, within the $\cN=2$ superconformal multiplet calculus, every minimal version of $\cN=2$ Poincar\'e or anti-de Sitter supergravity is realised as a locally superconformal coupling of the Weyl multiplet with two compensating multiplets, one of which is the Abelian vector multiplet.\footnote{There exists a version of 
 $\cN=2$ anti-de Sitter supergravity with a single compensator described by a massive tensor multiplet \cite{K-08, Butter:2010jm}. However, a compensating vector multiplet may be incorporated in this theory via a Stueckelberg-like mechanism.} 

In the present paper we propose an ${\cal N}=3$ nonlinear multiplet coupled to conformal supergravity and use it to formulate the equations of motion for ${\cal N} = 3$ Poincar\'e supergravity. 
At the heart of our construction is the  formalism of $\cN=3$ conformal superspace\footnote{The off-shell $\cN=3$ superconformal higher-spin gauge multiplets were introduced in \cite{KRHS} within the framework of $\cN=3$ conformal superspace with vanishing super-Weyl tensor. Importantly, these multiplets contain finitely many fields.} \cite{N=3CSS} which may be viewed to be  a superspace counterpart of the $\cN=3$ superconformal multiplet calculus proposed 
by van Muiden and Van Proeyen \cite{vanMvanP} and further developed by Hegde, Mishra and Sahoo \cite{HS, HMS}.\footnote{The field content of the $\cN=3$ Weyl multiplet \cite{vanMvanP} was predicted by Fradkin and Tseytlin \cite{FT}.} 
The $\cN=3$ conformal superspace of \cite{N=3CSS} is a natural development of the conformal superspace approached pioneered by Butter in the $\cN=1$ and $\cN=2$ cases in four dimensions
\cite{ButterN=1, ButterN=2},\footnote{These and other formulations for $\cN=1$ and $\cN=2$ conformal supergravity  have recently been reviewed in \cite{Review1,Review2}.} 
and further extended to conformal supergravity theories in two \cite{Kuzenko:2022qnb},
three \cite{BKNT-M1, BKNT-M2, KNT-M},  five \cite{ BKNT-M15} and six \cite{BKNT} dimensions.

In general, the superconformal multiplet calculus in diverse dimensions, as reviewed, e.g., in \cite{FT, FVP, Lauria:2020rhc},
has proved to be powerful in formulating general two-derivative supergravity-matter systems 
and studying their dynamical properties. In our opinion, the superconformal setting becomes especially powerful within superspace formulations for supergravity, which: (i) provide remarkably compact expressions for general supergravity-matter actions; (ii) make manifest the geometric properties of such theories; and
(iii) offer unique tools to generate higher-derivative couplings (of arbitrary order in fields and derivatives) in matter-coupled supergravity (see, e.g., 
\cite{Butter:2010jm, Butter:2013lta, Butter:2018wss}).
Superspace approaches become indispensable if one has to deal with off-shell 
supermultiplets with infinitely many auxiliary fields, including the following: 
(i) the charged $\cN=2$ hypermultiplet \cite{GIKOS, GIOS, LR1, LR2}; and (ii) the $\cN=3$ super Yang-Mills theory \cite{GIKOS1,GIKOS2,RoslyS}.

This paper is organised as follows. In section \ref{Section2} we briefly review $\cN=2$ conformal superspace and its degauged versions, the so-called $\sU(2)$ and $\sSU(2)$ superspaces, following \cite{ButterN=2}.
We also elaborate on the $\cN=2$ supergeometry with the Lorentz group being its structure group (no $R$-symmetry subgroup) which is naturally associated with the $\cN=2$ nonlinear multiplet, as described in subsection \ref{N=2ESS}. This supergeometry was called ``$\cN=2$ Einstein superspace'' in \cite{ButterHSS}. 
It was  originally introduced forty five years ago \cite{Wess:1979pp, Castellani:1980cu, Gates, Howe}. An important feature of $\cN=2$ Einstein superspace is that it does not allow maximally supersymmetric AdS backgrounds. 
This superspace was also originally used to develop the harmonic superspace formulation for $\cN=2$ supergravity
\cite{Galperin:1987em}. The latter was extended to conformal superspace by Butter in 2016  \cite{ButterHSS}. 

In section \ref{Section3} we review the $\cN=3$ conformal superspace and its degauged versions, following 
\cite{N=3CSS}. In section \ref{Section4} we introduce the $\cN=3$ nonlinear multiplet and apply it to construct 
$\cN=3$ Einstein superspace, which is characterised by the Lorentz group being its structure group. A remarkable feature of this supergeometry is that it allows a unique maximally supersymmetric background which is $\cN=3$ Minkowski superspace. We also describe the equations of motion for $\cN=3$ Poincar\'e supergravity. 
In section \ref{Section5} we describe some generalisations of the results. 

The main body of this paper is accompanied two appendices. In appendix \ref{AppendixA} 
our conventions are spelled out for the $\cN$-extended superconformal algebra of Minkowski superspace  $\mathfrak{su}(2,2|\cN)$. Appendix \ref{AppendixB} 
is devoted to a superspace description of the gauged $\cN=2$ supergravity introduced in \cite{deWvanHvanP}.

\section{$\cN=2$ curved supergeometries}
\label{Section2}

We begin this section by reviewing the $\cN=2$ conformal superspace formulation due to Butter \cite{ButterN=2}. Next, following \cite{ButterN=2} and \cite{ButterHSS}, we describe how, by performing a series of degaugings, one may reach all known $\cN=2$ curved superspace formulations. Specifically, we degauge to the: (i) $\sU(2)$ \cite{Howe}; (ii) $\sSU(2)$ \cite{Grimm,KLRT-M1}; and (iii) Einstein \cite{Wess:1979pp, Castellani:1980cu, Gates, Howe} superspaces. As an application of this dictionary, we describe $\cN=2$ Poincar\'e supergravity within superspace.

\subsection{$\cN=2$ conformal superspace}
\label{Section2.1}

Conformal superspace is a gauge theory of the superconformal algebra. It can be identified with a pair $(\cM^{4|8}, \nabla)$, where $\mathcal{M}^{4|8}$ denotes a supermanifold parametrised by local  coordinates\footnote{The Grassmann variables $\q^{\mu}_i $ and $\teb_{\dot{\mu}}^i$ are related to each other by complex conjugation: $\overline{\q^{\mu}_i}=\teb^{\dot{\mu}i}$.} $z^M = (x^m, \q^\m_\imath, \bar \q_{\dot \m}^\imath)$, where $m=0,1,\cdots,3$, $\mu=1,2$, $\dot{\mu}=1,2$ and  $i=\1,\2$, while $\nabla$ is a covariant derivative associated with the superconformal algebra. The generators $X_{\tilde A}$ of this algebra are described in appendix \ref{AppendixA}. They can be grouped in two disjoint subsets,
\bea 
X_{\tilde A} = (P_A, X_{\underline{A}} )~, \qquad
X_{\underline{A}} = (M_{ab} , \mathbb{Y}, {\mathbb J}_{ij}, {\mathbb D} , K^A)~,
\eea  
each of which constitutes a superalgebra:
\bsubeq\label{RigidAlgebra}
\begin{align}
	[P_{ {A}} , P_{ {B}} \} &= -f_{{ {A}} { {B}}}{}^{{ {C}}} P_{ {C}}
	\ , \\
	[X_{\underline{A}} , X_{\underline{B}} \} &= -f_{\underline{A} \underline{B}}{}^{\underline{C}} X_{\underline{C}} \ , \\
	[X_{\underline{A}} , P_{{B}} \} &= -f_{\underline{A} { {B}}}{}^{\underline{C}} X_{\underline{C}}
	- f_{\underline{A} { {B}}}{}^{ {C}} P_{ {C}}
	\ . \label{mixing}
\end{align}
\esubeq

In order to define the covariant derivatives, we associate with each generator $X_{\underline{A}}$ a connection one-form $\Omega^{\underline{A}}$
\begin{align}
	X_{\underline A} = (M_{ab} , \mathbb{Y}, {\mathbb J}_{ij}, {\mathbb D} , K^A) \quad \longleftrightarrow \quad (\O^{ab},\Phi,\F^{ij},B,\mathfrak{F}_{A}) = \O^{\underline{A}} = \rd z^M \O_M{}^{\underline A}~,
\end{align}
and with $P_{ {A}}$ a supervielbein one-form
$E^{ {A}} = \rd z^{ {M}} E_M{}^A$. 
It is assumed that the supermatrix $E_M{}^A$ is nonsingular 
\begin{align}
	E:= {\rm Ber} (E_M{}^A) \equiv {\rm sdet} (E_M{}^A)\neq 0~,
\end{align} 
hence there exists a unique inverse supervielbein. The latter is given by 
the supervector fields $E_A = E_A{}^M (z)\pa_M $, with 
$\pa_M = \pa /\pa z^M$, which constitute a new basis for the tangent space at each
point $z^M \in \cM^{4|8}$. The supermatrices $E_A{}^M $ and $E_M{}^A$ satisfy the 
properties:
\begin{align}
	E_A{}^ME_M{}^B=\d_A{}^B~, \qquad E_M{}^AE_A{}^N=\d_M{}^N~.
\end{align}
With respect to  the basis $E^A$,  the connection is expressed as 
$\Omega^{\underline{A}} =E^B\Omega_B{}^{\underline{A}}$, 
where $\Omega_B{}^{\underline{A}}=E_B{}^M\Omega_M{}^{\underline{A}}$. 
In this basis, the {covariant derivatives} $\nabla_A= (\nabla_a, \nabla_\a^i, \bar \nabla^\ad_i)$ are then given by 
\bea\label{nablaA}
\nabla_A 
&=& E_A  - \O_A{}^{\underline B} X_{\underline B}=
E_A -  \hf \Omega_A{}^{bc} M_{bc} - \ri \Phi_A {\mathbb Y} - \F_A{}^{jk} {\mathbb J}_{jk} - B_A \mathbb{D}
- \mathfrak{F}_{AB} K^B~.~~~
\eea
It should be noted that the translation generators $P_A$ do not appear above. Instead, the operators $\nabla_A$ replace $P_A$ and obey the graded commutation relations
\be
\label{3.2}
[ X_{\underline{B}} , \nabla_A \} = -f_{\underline{B} A}{}^C \nabla_C
- f_{\underline{B} A}{}^{\underline{C}} X_{\underline{C}} \ ,
\ee
compare with eq. \eqref{mixing}.

By definition, the gauge group of conformal supergravity is generated by local transformations of the form
\begin{subequations}\label{SUGRAtransmations}
	\bea
	\delta_\mathfrak{K} \nabla_A &=& [\mathfrak{K},\nabla_A] \ , \\
	\mathfrak{K} &=& \xi^B \nabla_B +  \L^{\underline{B}} X_{\underline{B}} ~, \non \\
	&=&  \xi^B \nabla_B+ \hf K^{bc} M_{bc} + \S \mathbb{D} + \ri \rho \mathbb{Y} 
	+ \theta^{jk} {\mathbb J}_{jk}
	+ \L_B K^B \ ,
	\eea
\end{subequations}
where  the gauge parameters satisfy natural reality conditions. Given a conformal tensor superfield $\mathfrak{T}$ (with indices suppressed), it transforms as 
\bea 
\label{3.14}
\d_{\mathfrak{K}} \mathfrak{T} = \mathfrak{K} \mathfrak{T} ~.
\eea
Such a superfield is said to be primary if it is 
characterised by the properties 
\bea
K^A  \mathfrak{T} = 0~, \quad \mathbb D \mathfrak{T} = \D \mathfrak{T} ~,  \quad {\mathbb Y} \mathfrak{T} = q \mathfrak{T}~,
\eea
for some real constants $\D$ and $q$, 
which are called the dimension and $\sU(1)_R$ charge of $\mathfrak{T}$ respectively.

This framework 
defines a geometric set-up to obtain a multiplet of conformal supergravity containing the conformal spin-2 field. However, in general, the resulting multiplet is reducible. To obtain an irreducible multiplet 
it is necessary to impose constraints on the algebra of covariant derivatives $[ \nabla_A , \nabla_B \}$. A beautiful feature of conformal superspace is the simplicity of the constraints needed to obtain the Weyl multiplet of conformal supergravity. In fact, to obtain a sufficient set of constraints, one requires this algebra to have a super Yang-Mills structure (compare with \cite{GSW})
\bsubeq \label{CSSAlgebra-0}
\bea
\{ \nabla_{\a}^i , \nabla_{\b}^j \}  &=&  -2 \ve^{ij} \ve_{\a \b} \bar{\cW} ~, \quad \{ \bar{\nabla}^{\ad}_i , \bar{\nabla}^{\bd}_j \} = 2 \ve_{ij} \ve^{\ad \bd} \cW ~, \\ &&\quad \;\; \{\nabla_{\a}^i , \bar{\nabla}^{\bd}_{j} \} = - 2 \ri \d^i_j \nabla_{\a}{}^{\bd} ~.
\eea
\esubeq
Here the operator $\bar{\cal{W}}$ is the complex conjugate of $\cal{W}$ in the sense that $\overline{\cW U} = \bar{\cW} \bar U$, for any tensor superfield. The operator $\cW$ takes the form
\bea
{\cal{W}}
&=&
\frac{1}{2}{\cal{W}}(M)^{ab} M_{ab}
+\ri {\cal{W}}(\mathbb Y)\mathbb Y
+\cW(J)^{ij} J_{ij}
+{\cal{W}}(\mathbb D)\mathbb D
\non\\
&&
+{\cal{W}}(S)_\a^i S^\a_i
+{\cal{W}}(\bar S)_{i}^\ad \bar{S}_\ad^i
+{\cal{W}}(K)_a K^a
~.
\eea
Having imposed the constraints \eqref{CSSAlgebra-0}, the graded Jacobi identities
\be
(-1)^{\varepsilon_{{A}}  \varepsilon_{{C}}}[\nabla_{A}, [\nabla_{B}, \nabla_{C} \} \} 
~+~ \text{(two cycles)}
= 0 \ ,
\label{Jacobi-0}
\ee
 become non-trivial and play the role of consistency conditions which may be used to determine the torsion and curvature. Their solution is as follows:
\begin{subequations}
	\label{CSSAlgebra}
	\begin{align}
		\{ \nabla_\a^i , \nabla_\b^j \} &= 2 \ve^{ij} \ve_{\a\b} \big( \bar{W}_{\gd\dd} \bar{M}^{\gd\dd} + \frac 1 4 \bar{\nabla}_{\gd k} \bar{W}^{\gd\dd} \bar{S}^k_\dd - \frac 1 4 \nabla_{\g\dd} \bar{W}^\dd{}_\gd K^{\g \gd} \big)~, \\
		\{ \bar{\nabla}^\ad_i , \bar{\nabla}^\bd_j \} &= -2 \ve_{ij} \ve^{\ad\bd} \big( W^{\g\d} M_{\g\d} - \frac 1 4 \nabla^{\gd k} W_{\g\d} S_k^\d + \frac 1 4 \nabla^{\g\gd} W_\g^\d K_{\d \gd} \big)~, \\
		\{ \nabla_\a^i , \bar{\nabla}^\bd_j \} &= - 2 \ri \d_j^i \nabla_\a{}^\bd~.
	\end{align}
\end{subequations}

Remarkably, this algebra is expressed in terms of a single superfield $W_{\alpha \beta }= W_{(\a\b)}$, its conjugate $\bar W_{\ad \bd}$, and their covariant derivatives. This superfield is an $\mathcal{N}=2$ extension of the Weyl tensor, and is called the super-Weyl tensor. It proves to be a primary chiral superfield of dimension 1,
\bea
K^C W_{\a \b } =0~, \quad \bar \nabla^\gd_k W_{\a\b}=0 ~, \quad
{\mathbb D} W_{\a\b} = W_{\a\b}~,\quad
{\mathbb Y} W_{\a\b} = -2 W_{\a\b}~,
\label{constr-W-1}
\eea
and it obeys the Bianchi identity
\bsubeq\label{N=2Bach}
\bea
B&:=&\nabla_{\a \b} W^{\a\b}
= \bar \nabla^{\ad \bd} \bar W_{\ad \bd}=B~,
\label{super-Bach} \\
\nabla_{\a \b} &:=& \nabla_{(\a}^{i} \nabla_{\b) i} ~, \qquad \bar{\nabla}^{\ad \bd} := \bar{\nabla}^{(\ad}_i \bar{\nabla}^{\bd) i} ~,
\eea
\esubeq
where $B$ is the $\cN=2$ super-Bach scalar.\footnote{At the component level it contains the Bach tensor.} Setting $B=0$ gives the equation of motion for $\cN=2$ conformal supergravity.

\subsection{Degauging (i): $\sU(2)$ superspace} \label{subsection2.2}

According to \eqref{SUGRAtransmations}, under an infinitesimal special superconformal gauge transformation, the dilatation connection transforms as follows
\bea
\mathfrak{K} = \Lambda_{B} K^{B} \quad \implies \quad
\d_{\mathfrak{K}} B_{A} = - 2 \Lambda_{A} ~.
\eea
Thus, it is possible to choose the gauge condition
$B_{A} = 0$, which completely fixes 
the special superconformal gauge freedom.\footnote{There is a class of residual gauge transformations preserving the gauge $B_{A}=0$. These generate the super-Weyl transformations of $\sU(2)$ superspace, see the next subsection.} As a result, the corresponding connection is no longer required for the covariance of $\nabla_A$ under the residual gauge freedom and
should be separated,
\bea
\nabla_{A} &=& \mathfrak{D}_{A} - \mathfrak{F}_{AB} K^{B} ~. \label{ND}
\eea
Here the operator $\mathfrak{D}_{A} $ involves only the Lorentz and $\sU(2)_R$ connections
\bea
\mathfrak{D}_A = E_A - \frac{1}{2} \O_A{}^{bc} M_{bc} - \ri \F_A \mathbb{Y} - \F_A{}^{jk} \mathbb{J}_{jk} ~.
\eea

The next step is to relate the special superconformal connection
$\mathfrak{F}_{AB}$  to the torsion tensor of $\sU(2)$ superspace. To do this, one can  make use of the relation
\bea
\label{4.3}
[ \nabla_{A} , \nabla_{B} \} &=& [ \mathfrak{D}_{A} , \mathfrak{D}_{B} \} - \big(\mathfrak{D}_{A} \mathfrak{F}_{BC} - (-1)^{\epsilon_A \epsilon_B} \mathfrak{D}_{B} \mathfrak{F}_{AC} \big) K^C - \mathfrak{F}_{AC} [ K^{C} , \nabla_B \} \non \\
&& + (-1)^{\epsilon_A \epsilon_B} \mathfrak{F}_{BC} [ K^{C} , \nabla_A \} + (-1)^{\epsilon_B \epsilon_C} \mathfrak{F}_{AC} \mathfrak{F}_{BD} [K^D , K^C \} ~.
\eea
In conjunction with \eqref{CSSAlgebra}, this relation leads to a set of consistency conditions that are equivalent to the Bianchi identities of $\sU(2)$ superspace \cite{Howe}. 
Their solution expresses the components of $\mathfrak{F}_{AB}$ in terms of the torsion 
tensor of $\sU(2)$ superspace and completely determines the geometry of the $\mathfrak{D}_{A}$ derivatives \cite{ButterN=2}. 
The outcome of the analysis is as follows:
\begin{subequations} \label{connections}
	\bea
	\mathfrak{F}_\a^i{}_\b^j  
	&=&
	-\hf\ve_{\a\b}S^{ij}
	+\hf\ve^{ij}Y_{\a\b}
	~,
	\\
	\mathfrak{F}^\ad_i{}^\bd_j  
	&=&
	-\hf\ve^{\ad\bd}\bar{S}_{ij}
	+\hf\ve_{ij}\bar{Y}^{\ad\bd}
	~,\\
	\mathfrak{F}_\a^i{}^\bd_j
	&=&
	- \mathfrak{F}^\bd_j{}_\a^i
	=
	-\d^i_jG_\a{}^\bd
	-\ri G_\a{}^\bd{}^i{}_j
	~.
	\eea
\end{subequations}
The dimension-1 superfields introduced above have the following symmetry properties:  
\bea
S^{ij}=S^{ji}~, \qquad Y_{\a\b}=Y_{\b\a}~, \qquad G_{\a\ad}{}^{ij}=G_{\a\ad}{}^{ji}~.
\eea
They obey the reality conditions:
\bea
\overline{S^{ij}} =  \bar{S}_{ij}~,\quad
\overline{Y_{\a\b}} = \bar{Y}_{\ad\bd}~,\quad
\overline{G_{\b\ad}} = G_{\a\bd}~,\quad
\overline{G_{\b\ad}{}^{ij}} = ~G_{\a\bd}{}_{ij}
~.~~~~~~
\eea
The ${\sU}(1)_R$ charges of the complex fields are:
\bea
{\mathbb Y} \,S^{ij}=2S^{ij}~,\qquad
{\mathbb Y}  \,Y_{\a\b}=2Y_{\a\b}~.
\eea
The algebra obeyed by ${\mathfrak{D}}_A$ takes the form:
\begin{subequations} \label{U(2)algebra}
	\bea
	\{ \mathfrak{D}_\a^i , \mathfrak{D}_\b^j \}
	&=&
	4 S^{ij}  M_{\a\b} 
	+2\ve_{\a\b}\ve^{ij}Y^{\g\d}  M_{\g\d}  
	+2\ve^{ij} \ve_{\a\b}  \bar{W}_{\gd\dd} \bar{M}^{\gd\dd} 
	\non\\
	&&
	+2\ve_{\a\b}\ve^{ij}S^{kl} \mathbb{J}_{kl}
	+ 4Y_{\a\b}  \mathbb{J}^{ij}
	~,
	\label{U(2)algebra.a}
	\\
	\{ \mathfrak{D}_\a^i , \bar{\mathfrak{D}}^\bd_j \}
	&=&
	- 2 \ri \d_j^i\mathfrak{D}_\a{}^\bd
	+4\Big(
	\d^i_jG^{\g\bd}
	+\ri G^{\g\bd}{}^i{}_j
	\Big) 
	M_{\a\g} 
	+4\Big(
	\d^i_jG_{\a\gd}
	+\ri G_{\a\gd}{}^i{}_j
	\Big)  
	\bar{M}^{\bd\gd}
	\non\\
	&&
	+8 G_\a{}^\bd \mathbb{J}^i{}_j
	-4\ri\d^i_j G_\a{}^\bd{}^{kl} \mathbb{J}_{kl}
	-2\Big(
	\d^i_jG_\a{}^\bd
	+\ri G_\a{}^\bd{}^i{}_j
	\Big)
	\mathbb{Y} 
	~.
	\label{U(2)algebra.b}
	\eea
	\esubeq
	The consistency conditions arising from solving eq. \eqref{4.3} and the constraints satisfied by $W_{\a\b}$ in conformal superspace lead to the following set of dimension-3/2 Bianchi identities:
	\begin{subequations}\label{BI-U2}
		\bea
		\bar{\mathfrak{D}}_{\ad i} {W}_{\b\g}&=&0~,
		\\
		\mathfrak{D}_{\a}^{(i}S^{jk)}&=&0~,
		\\
		\mathfrak{D}_{(\a}^{i}Y_{\b\g)}&=&0~,
		\\
		\mathfrak{D}_{(\a}^{(i}G_{\b)\bd}{}^{jk)}&=&0~, 
		\\
		\bar{\mathfrak{D}}_{\ad}^{(i}S^{jk)} &=& \ri\mathfrak{D}^{\b (i}G_{\b\ad}{}^{jk)}~,
		\\
		\mathfrak{D}_{\a}^{i}S_{ij}
		&=&
		-\mathfrak{D}^{\b}_{j}Y_{\b\a}
		~,
		\\
		\mathfrak{D}_\a^iG_{\b\bd}&=&
		- \frac{1}{ 4}\bar{\mathfrak{D}}_\bd^iY_{\a\b}
		+ \frac{1}{ 12}\ve_{\a\b}\bar{\mathfrak{D}}_{\bd j}S^{ij}
		- \frac{1}{ 4}\ve_{\a\b}\bar{\mathfrak{D}}^{\gd i}\bar{W}_{\gd\bd}
		- \frac{\ri }{ 3}\ve_{\a\b}\mathfrak{D}^{\g}_j G_{\g \bd}{}^{ij}
		~.
		\eea
	\end{subequations}
	Here we have used the definitions
	\begin{align}
		\mathfrak{D}_{\a \b} = \mathfrak{D}^i_{(\a} \mathfrak{D}_{\b) i} ~, \qquad \bar{\mathfrak{D}}_{\ad \bd} = \bar{\mathfrak{D}}_i^{( \ad } \bar{\mathfrak{D}}^{\bd ) i}~.
	\end{align}
	It is useful to also define
	\begin{align}
		\mathfrak{D}^{ij} = \mathfrak{D}^{\a (i} \mathfrak{D}_{\a}^{j)} ~, \qquad \bar{\mathfrak{D}}_{i j} = \bar{\mathfrak{D}}_{\ad (i} \bar{\mathfrak{D}}^{\ad}_{j)}~.
	\end{align}
	
	As is well known, the gauge freedom of $\sU(2)$ superspace includes both local structure group and super-Weyl transformations. These correspond to the residual gauge transformations of \eqref{SUGRAtransmations} in the gauge $B_A=0$. In particular, it is clear that this includes local $\mathscr{K}$ transformations of the form 
		\bea
		\label{U(2)StructureGP}
		\delta_\mathscr{K} \mathfrak{D}_A = [\mathscr{K},\mathfrak{D}_A] ~, \qquad \mathscr{K} = \xi^B \mathfrak{D}_B + \hf K^{bc} M_{bc} + \theta^{jk} {\mathbb J}_{jk} + \ri \rho \mathbb{Y} ~.
		\eea
	
	These transformations are not the most general conformal supergravity gauge transformations preserving the gauge $B_A=0$. Specifically, it may be shown that the following transformation also enjoys this property
	\begin{align}
		\mathfrak{K}(\Sigma) =  \S \mathbb{D} + \hf \nabla_B \S K^{B} \quad \implies \quad \d_{\mathfrak{K}( \S )} B_A = 0~.
	\end{align}
	Now, by making use of the identity
	\bea
	\nabla'_{A} = \mathfrak{D}'_{A} - \mathfrak{F}'_{AB} K^{B} = \re^{\mathfrak{K}(\Sigma)}  \nabla_{A} \re^{-\mathfrak{K}(\Sigma)} ~,
	\eea
	we may determine how finite transformations of this form manifest in $\sU(2)$ superspace. One finds that the covariant derivatives transform as:
	\begin{subequations}
	\bea
	\mathfrak{D}^{' i}_\a&=&\re^{\hf \S} \Big( \mathfrak{D}_\a^i+2\mathfrak{D}^{\b i}\S M_{\a \b} + 2 \mathfrak{D}_{\a}^j \S \mathbb{J}^{j}{}_i 
	- \frac 1 2 \mathfrak{D}_\a^i\S {\mathbb Y} \Big)
	~,
	\\
	\bar{\mathfrak{D}}_{i}^{' \ad}&=&\re^{\hf \S} \Big( \bar{\mathfrak{D}}^\ad_i+2 \bar{\mathfrak{D}}_{i}^{\bd} \S \bar{M}_{\ad \bd} - 2 \bar{\mathfrak{D}}^{\ad}_j \S \mathbb{J}^{j}{}_i 
	+ \frac{1}{2} \bar{\mathfrak{D}}_{i}^{\ad} \S {\mathbb Y} \Big)
	~, \\
	{\mathfrak{D}}'_\aa&=&\re^{\S} \Big( \mathfrak{D}_{\a \ad} + {\rm i} \mathfrak{D}^i_{\a} \S \cDB_{\ad i} + {\rm i} \bar{\mathfrak{D}}_{\ad i} \S \mathfrak{D}_{\a}^i + \Big( \mathfrak{D}^\b{}_\ad \S - \ri \mathfrak{D}^{\b i} \S \bar{\mathfrak{D}}_{\ad i} \S \Big) M_{\a \b} \non \\ && + \Big( \mathfrak{D}_{\a}{}^\bd \S + \ri \mathfrak{D}_{\a}^i \S \bar{\mathfrak{D}}^{\bd}_i \S \Big) { \bar M}_{\ad \bd} 
	- 2\ri \mathfrak{D}_\a^i \S \bar{\mathfrak{D}}_{\ad j} \S \mathbb{J}^j{}_i
	+ \frac{\ri}{2} \mathfrak{D}_{\a}^i \S \bar{\mathfrak{D}}_{\ad i} \S \mathbb{Y} 
	\Big)
	~,
	\eea
	\end{subequations}
	while the corresponding transformations for the torsion superfields are:
	\begin{subequations}
		\bea
		S^{' ij}&=& \re^{\S} \Big( S^{ij}
		-\hf \cD^{ij} \S + \mathfrak{D}^{\a (i} \S \mathfrak{D}_\a^{j)} \S \Big)~,
		\\
		Y^{'}_{\a\b}&=& \re^{\S} \Big( Y_{\a\b}
		-\hf \mathfrak{D}_{\a \b} \S - \mathfrak{D}_\a^{i} \S \mathfrak{D}_{\b i} \S \Big)~,
		\\
		G'_{\a\ad}&=&
		\re^{\S} \Big( G_{\a\ad}
		-{\frac{1}{8}}[\mathfrak{D}_\a^i,\bar{\mathfrak{D}}_{\ad i}]\S
		-\frac{1}{2} \mathfrak{D}_\a^i \S \bar{\mathfrak{D}}_{\ad i} \S \Big)
		~,
		\\
		G'_{\a\ad}{}^{ij}&=& \re^{\S} \Big( G_{\a\ad}{}^{i}{}_j
		+{\frac\ri 4} [\mathfrak{D}_\a^{(i},\bar{\mathfrak{D}}_{\ad}^{j)}] \S \Big)
		~, \label{Finite_Gij}\\
		W'_{\a\b} &=& \re^{\S} W_{\a \b}~.
		\eea
	\end{subequations}
	These are exactly the super-Weyl transformations of $\sU(2)$ superspace.

\subsection{Degauging (ii): $\sSU(2)$ superspace}
\label{section2.3}
	
	The torsion $G_{\a\ad}{}^{ij}$ of $\sU(2)$ superspace proves to describe purely gauge degrees of freedom, see \cite{Howe,KLRT-M2}. Specifically, one can use super-Weyl gauge freedom \eqref{Finite_Gij} to set
	\bea
	G_{\a\bd}{}^{ij}=0~.
	\label{G2}
	\eea
	In this gauge, it is natural to  introduce new covariant derivatives $\cD_A$ defined by
	\bea
	\mathscr{D}_\a^i = \mathfrak{D}_\a^i~, \qquad
	\mathscr{D}_\aa=\mathfrak{D}_\aa-\ri \,G_\aa \,{\mathbb Y}~.
	\eea
	Making use of \eqref{U(2)algebra}, we find that they obey the algebra
	\begin{subequations} 
		\label{SU(2)algebra}
		\bea
		\{\mathscr{D}_\a^i,\mathscr{D}_\b^j\}&=&
		4S^{ij}M_{\a\b}
		+2\ve^{ij}\ve_{\a\b}Y^{\g\d}M_{\g\d}
		+2\ve^{ij}\ve_{\a\b}\bar{W}^{\gd\dd}\bar{M}_{\gd\dd}
		\non\\
		&&
		+2 \ve_{\a\b}\ve^{ij}S^{kl}J_{kl}
		+4 Y_{\a\b}J^{ij}~, 
		\\
		\{\mathscr{D}_\a^i,\bar{\mathscr{D}}^\bd_j\}&=&
		-2\ri\d^i_j\mathscr{D}_\a{}^{\bd}
		+4\d^{i}_{j}G^{\d\bd}M_{\a\d}
		+4\d^{i}_{j}G_{\a\gd}\bar{M}^{\gd\bd}
		+8 G_\a{}^\bd J^{i}{}_{j}~,~~~~~~~~~
		\eea
	\end{subequations}
	The various torsion tensors in \eqref{SU(2)algebra} obey the Bianchi identities \eqref{BI-U2} upon the replacement $\mathfrak{D}_A \rightarrow \mathscr{D}_A$ and imposing eq. \eqref{G2}. Examining equations \eqref{SU(2)algebra}, we see that the $\sU(1)_R$ curvature has been eliminated, hence corresponding connection is flat and may be gauged away by some local $\sU(1)_R$ transformation
	\be
	\label{4.23}
	\F_A=0~.
	\ee
	As a result, the gauge group reduces to $\sSL( 2, {\mathbb C}) \times \sSU(2)_R$ and the superspace geometry is the so-called $\sSU(2)$ superspace of \cite{Grimm}.

	It turns out that the gauge conditions \eqref{G2} and \eqref{4.23} allow for residual structure group and super-Weyl transformations.
	The former are defined analogously to their $\sU(2)$ superspace cousins \eqref{U(2)StructureGP}, with the defining difference that the $\sU(1)_R$ parameter is switched off; $\r = 0$.
	On the other hand, the residual super-Weyl transformations are described
	by a parameter $\S$ constrained by
	\be
	[\mathfrak{D}_\a^{(i},\bar{\mathfrak{D}}_\ad^{j)}]\S=0~.
	\label{Ucon}
	\ee
	The general solution of this condition is \cite{KLRT-M1}
	\bea
	\S = \frac{1}{2} (\s+\bar{\s})~, \qquad \bar{\mathfrak{D}}^\ad_i \s =0~,
	\qquad {\mathbb Y}\, \s =0~,
	\eea
	where the parameter $\s$ is covariantly chiral, with zero $\sU(1)_R$ charge, but otherwise arbitrary.
	To preserve the gauge condition $\F_A=0$, every super-Weyl transformation must be accompanied by the following compensating $\sU(1)_R$ transformation 
	\be
	\mathfrak{D}_A \longrightarrow \re^{- \frac 1 4 (\s - \bar{\s}) \mathbb{Y}} \mathfrak{D}_A \re^{ \frac 1 4 (\s - \bar{\s}) \mathbb{Y}}
	\ee
	As a result, the $\sSU(2)$ geometry is preserved by the following super-Weyl transformations \cite{KLRT-M1}:
\begin{subequations}
	\label{SU(2)SW}
	\begin{align}
		\mathscr{D}_\a^{'i}&= \re^{\frac 1 2 \bar{\s}} \Big( \mathscr{D}_\a^i+ \mathscr{D}^{\b i}\s M_{\a \b} + \mathscr{D}_{\a j}\s \mathbb{J}^{ij} \Big) ~, 
		\\ 
		\bar{\mathscr{D}}_{i}^{' \ad}&=\re^{\frac{1}{2} \s} \Big( \mathscr{D}^\ad_i-2 \bar{\mathscr{D}}_{ \bd i} \bar{\s} \bar{M}^{\ad \bd} - \bar{\mathscr{D}}^{\ad j} \bar{\s} \mathbb{J}_{ij} \Big)~,
		\\
		\mathscr{D}_\aa' &= \re^{\hf \s + \hf \bar{\s}} \Big(\mathscr{D}_\aa + \frac{\rm i}{2} \mathscr{D}^i_{\a} \s \bar{\mathscr{D}}_{\ad i} + \frac{\rm i}{2} \bar{\mathscr{D}}_{\ad i} \bar{\s} \mathscr{D}_{\a}^i + \hf \Big( \mathscr{D}^\b{}_\ad (\s + \bar \s ) - \frac{\ri}{2} \mathscr{D}^{\b i} \s \bar{\mathscr{D}}_{\ad i} \bar{\s} \Big) M_{\a \b} \non \\ & + \hf \Big( \mathscr{D}_{\a}{}^\bd (\s + \bar{\s}) + \frac{\ri}{2} \mathscr{D}_{\a}^i \s \bar{\mathscr{D}}^{\bd}_i \bar{\s} \Big) { \bar M}_{\ad \bd} 
		+ \frac \ri 2 \mathscr{D}_\a^i \s \bar{\mathscr{D}}_{\ad}^j \bar{\s} \mathbb{J}_{ij} \Big) ~, \\
		S^{' ij}&= \re^{\bar{\s}} \Big( S^{ij}-{\frac14}\mathscr{D}^{ij} \s + \frac 1 4 \mathscr{D}^{\a (i} \s \mathscr{D}_\a^{j)} \s \Big)~,  \\
		Y^{'}_{\a\b}&= \re^{\bar{\s}} \Big( Y_{\a\b}-{\frac14}\mathscr{D}_{\a \b}\s - \frac14 \mathscr{D}_\a^{i} \s \mathscr{D}_{\b i} \s \Big)~,\\
		G_{\aa}' &=
		\re^{\frac12 \s + \frac12 \bar{\s}} \Big(
		G_{\a\bd} -{\frac{\ri}4}
		\mathscr{D}_{\a \ad} (\s-\bar{\s})
		-\frac{1}{8} \mathscr{D}_\a^i \s \bar{\mathscr{D}}_{\ad i} \bar{\s}
		\Big)~, \non \\
		W'_{\a \b} &= \re^{\s} W_{\a \b}~.
	\end{align}
\end{subequations}
Here we have made use of the definitions
\begin{align}
	\mathscr{D}_{\a \b} &= \mathscr{D}^i_{(\a} \mathscr{D}_{\b) i} ~, \qquad \mathscr{D}^{ij} = \mathscr{D}^{\a (i} \mathscr{D}_{\a}^{j)}~, \non \\
	\bar{\mathscr{D}}^{\ad \bd} &= \bar{\mathscr{D}}_i^{( \ad }\bar{\mathscr{D}}^{\bd ) i} ~, \qquad \bar{\mathscr{D}}_{i j} = \bar{\mathscr{D}}_{\ad (i}\bar{\mathscr{D}}^{\ad}_{j)}~.
\end{align}

\subsection{Degauging (iii): $\cN = 2$ Einstein superspace}
\label{N=2ESS}

In the framework of conformal superspace, the nonlinear multiplet is described by a matrix superfield $L_{\bold a}^i$, where $i$ is an $\sSU(2)_R$ index while $\bold{a}$ corresponds to an external $\sSU(2)$ group. It satisfies the reality condition $\overline{L_{\bold a}^i} = - L^{\bold a}_i$, and is characterised by the nonlinear constraints
\begin{align}
	\label{SU(2)NL}
	\text{det}\, L = 1 ~, \qquad L^{\bold a (i} \nabla^{j}_\a L^{k)}_{\bold a} = 0~,
\end{align}
which are superconformal provided that $L_{\bold a}^i$ is primary, dimensionless, and uncharged. These constraints may be trivially degauged to the $\sSU(2)$ superspace described above by making the replacement $\nabla \rightarrow \mathscr{D}$ in eq. \eqref{SU(2)NL}.

The nonlinear multiplet first arose within the component construction of \cite{deWvanHvanP}, where it was used as one of the two  compensating multiplets.
Specifically, it was utilised to break the $\sSU(2)_R$ invariance of conformal supergravity. This can be achieved at the $\sSU(2)$ superspace level by requiring
\begin{align}
	\label{SU(2)gauge}
	L_{\bold a}^i = \d_{\bold{a}}^i~,
\end{align}
see e.g. \cite{ButterHSS}.
Since this constraint fixes the $\sSU(2)_R$ freedom, the corresponding connection in $\mathscr{D}_A$ should be separated
\begin{align}
	\mathscr{D}_A = \cD_A + \Phi_A{}^{jk} {\mathbb J}_{jk}~, \qquad \cD_A = E_A{}^M \partial_M + \hf \O_A{}^{bc} M_{bc}~.
\end{align}
Consistency with the differential constraint in eq. \eqref{SU(2)NL} leads to the following condition on the degauged $\sSU(2)$ connection
\begin{align}
	\label{2.16}
	\Phi_\a^{(i, jk)} = 0~,
\end{align}
which implies
\begin{align}
	\mathscr{D}_\a^i = \cD_\a^i - \chi_{\a j} \mathbb{J}^{ij}~,
\end{align}
where $\chi_\a^i$ is a dimension-$1/2$ superfield. This allows us to determine the geometry described by $\cD_A$. Specifically, we see that:
\begin{subequations}
	\begin{align}
		\{ \cD_{(\a}^{(i} , \cD_{\b)}^{j)} \} &= \{ \mathscr{D}_{(\a}^{(i} , \mathscr{D}_{\b)}^{j)}\} + \chi_{(\a}^{(i} \cD_{\b)}^{j)} + 2 \cD_{(\a}^{(i} \chi_{\b) k} {\mathbb J}^{j)k} + \hf \chi_\a^k \chi_{\b k} {\mathbb J}^{ij}  ~, \\
		\{ \cD^{\a i} , \cD_{\a i} \} &= \{ \mathscr{D}^{\a i} , \mathscr{D}_{\a i}\} + 3 \chi^\a_i \cD_\a^i - 2 \cD^\a_i \chi_{\a j} {\mathbb J}^{ij} - \chi^\a_i \chi_{\a j} {\mathbb J}^{ij}~, \\
		\{ \cD_{\a}^{(i} , \cDB_{\ad}^{j)} \} &= \{ \mathscr{D}_{\a}^{(i} , \bar{\mathscr{D}}_{\ad}^{j)} \} + \hf \chi_\a^{(i} \cDB_\ad^{j)} + \hf \bar{\chi}_\ad^{(i} \cD_\a^{j)} + ( \cDB_\ad^{(i} \chi_{\a k} + \cD_\a^{(i} \bar{\chi}_{\ad k}) {\mathbb J}^{j)k} + \hf \chi^{k}_\a \bar{\chi}_{\ad k} {\mathbb J}^{ij}~, \\
		\{ \cD_{\a}^{i} , \cDB_{\ad i} \} &= \{ \mathscr{D}_{\a}^{i} , \bar{\mathscr{D}}_{\ad i} \} - \frac 3 2 \bar{\chi}_\ad^i \cD_{\a i} - \frac{3}{2} \chi_{\a i} \bar{\cD}_\ad^i - (\cD_\a^i \bar{\chi}_\ad^j - \bar{\cD}_\ad^i \chi_\a^j - \chi_\a^i \bar{\chi}_\ad^j) {\mathbb J}_{ij}~.
	\end{align}
\end{subequations}

Owing to eq. \eqref{2.16}, it is clear that the $\sSU(2)_R$ curvature in each equation above identically vanishes. This has highly nontrivial consequences. In particular, we find that the torsions $S^{ij}$, $Y_{\a \b}$ and $G_\aa$ of $\sSU(2)$ superspace become descendants of the spinor torsion $\chi_\a^i$:
\begin{subequations}
	\begin{align}
		S^{ij} &= - \frac 1 4 \cD^{\a (i} \chi_\a^{j)} - \frac 1 8 \chi^{\a (i} \chi_\a^{j)}~, \\
		Y_{\a \b} &= - \frac 1 4 \cD_{(\a}^i \chi_{\b) i} - \frac 1 8 \chi_{(\a}^i \chi_{\b) i}~, \\
		G_{\aa} &= \frac 1 {16} \Big( \cDB_{\ad i} \chi_\a^i - \cD_\a^i \bar{\chi}_{\ad i} - \chi_\a^i \bar{\chi}_{\ad i} \Big)~.
	\end{align}
\end{subequations} 
This implies that the resulting geometry is described solely in terms of $\chi_\a^i$ and the super-Weyl tensor $W_{\a \b}$, which obey the Bianchi identities
\begin{subequations}
\begin{align}
		\cDB_\ad^i W_{\a \b}  &= 0 ~, \qquad \cD^\b_i W_{\a \b} + 2 \cDB^\ad_i G_{\aa} - \frac{1}{3} (\cD_{\a}^j + 2 \c_{\a}^j) \bar{S}_{ij} = 0 ~, \\
		\cD_{(\a}^{(i} \chi_{\b)}^{j)} &= 0 ~, \qquad \cD_\a^{(i} \bar{\chi}_\ad^{j)} - \frac 1 2 \chi_\a^{(i} \bar{\chi}_\ad^{j)} = 0~.
\end{align}
\end{subequations}
As a result, we find the algebra of degauged covariant derivatives $\cD_A$ to be \cite{Howe, Castellani:1980cu, Gates}:
\begin{subequations} \label{N=2EinsteinGeometry}
	\begin{align}
		\{ \cD_\a^i , \cD_\b^j \} &= \chi_{(\a}^{(i} \cD_{\b)}^{j)} - \frac 3 4 \ve_{\a\b} \ve^{ij} \chi^\g_k \cD_\g^k + 4 S^{ij} M_{\a \b} + 2 \ve_{\a \b} \ve^{i j} \Big ( Y^{\g \d} M_{\g \d} + \bar{W}^{\gd \dd} \bar{M}_{\gd \dd} \Big ) ~, \\
		\{ \cD_\a^i , \cDB^\ad_j \} &= \frac 1 2 \Big( \bar{\chi}^\ad_j \cD_\a^i - 2 \d^i_j \bar{\chi}^\ad_k \cD_\a^k \Big) + \frac 1 2 \Big( \chi_\a^i \cDB^\ad_j - 2 \d^i_j \chi_\a^k \cDB^\ad_k \Big) - 2 \ri \d^i_j \cD_\a{}^{\ad} \non \\
		&\phantom{=} + 4 \d^i_j \Big( G^{\b \ad} M_{\a \b} + G_{\a \bd} \bar{M}^{\ad \bd} \Big)~.
	\end{align}
\end{subequations}

The supergravity gauge freedom of this geometry includes both local $\cK$ and super-Weyl transformations. The former transform $\cD_A$ as follows
\begin{align}
	\d_\cK \cD_A = [\cK, \cD_A] ~, \qquad \cK = \xi^B \cD_B + \frac 1 2 K^{ab} M_{ab}~,
\end{align}
and act on tensor superfields $\cT$ as $\d_\cK \cT = \cK \cT$. Keeping in mind equations \eqref{SU(2)SW} and \eqref{2.16}, it is clear that the super-Weyl transformations of this supergeometry take the form
\bsubeq
\bea
\mathcal{D}_\a^{'i}&=& \re^{\frac 1 2 \bar{\s}} \Big( \mathcal{D}_\a^i+ \mathcal{D}^{\b i}\s M_{\a \b} \Big) ~, 
\\ 
\bar{\mathcal{D}}_{i}^{' \ad}&=&\re^{\frac{1}{2} \s} \Big( \mathcal{D}^\ad_i- \bar{\mathcal{D}}_{ \bd i} \bar{\s} \bar{M}^{\ad \bd} \Big)~,
\\
\mathcal{D}_\aa' &=& \re^{\hf \s + \hf \bar{\s}} \Big(\mathcal{D}_\aa + \frac{\rm i}{2} \mathcal{D}^i_{\a} \s \bar{\mathcal{D}}_{\ad i} + \frac{\rm i}{2} \bar{\mathcal{D}}_{\ad i} \bar{\s} \mathcal{D}_{\a}^i + \hf \Big( \mathcal{D}^\b{}_\ad (\s + \bar \s ) - \frac{\ri}{2} \mathcal{D}^{\b i} \s \bar{\mathcal{D}}_{\ad i} \bar{\s} \Big) M_{\a \b} \non \\ && + \hf \Big( \mathcal{D}_{\a}{}^\bd (\s + \bar{\s}) + \frac{\ri}{2} \mathcal{D}_{\a}^i \s \bar{\mathcal{D}}^{\bd}_i \bar{\s} \Big) { \bar M}_{\ad \bd} \Big) ~, \\
\chi_\a^{'i} &=& \re^{\frac 1 2 \bar{\s}} \Big( \chi_\a^i + \cD_\a^i \s \Big)~, \\
{W}'_{\a \b}&=& \re^{\s} {W}_{\a \b }~,
\eea
\end{subequations}
where $\s$ is a chiral parameter, $\bar{\cD}_i^\ad \s = 0$.

\subsection{$\cN=2$ Poincar\'e supergravity}
\label{section2.5}

To conclude this section, we provide an application of the superspace formulations reviewed above. Specifically, we will utilise them to describe Poincar\'e supergravity. We recall that the latter may described by coupling the Weyl multiplet of conformal supergravity to two compensators, a vector multiplet and nonlinear multiplet, which can be used reduce the local structure group to the Lorentz group.

Thus, we begin by coupling the Weyl multiplet to a vector multiplet $\cW$ and nonlinear multiplet $L_{\bold a}^i$. The former is a reduced chiral superfield
\begin{align}
	\bar{\nabla}^\ad_i \cW = 0 ~, \qquad \nabla^{ij} \cW = \bar{\nabla}^{ij} \bar{\cW} ~,
\end{align}
while the latter is subject to the constraints \eqref{SU(2)NL}. 
The supergravity equations of motion are described by 
the primary constraints
\begin{subequations}\label{PoincareEoM}
	\bea
	 \nabla^{ij}  \cW &=& 0~, \label{PoincareEoM.a}\\
	 L^{\bold a}_j \nabla_\a^j L_{\bold a}^i + \frac{3}{2} \nabla_\a^i \log \cW &= &0~. \label{PoincareEoM.b}
	\eea
\end{subequations}
Equation \eqref{PoincareEoM.a} is the equation of motion for the vector multiplet derived from the action 
\bea
S_{\rm SG}
&=&- \frac{1}{ 4\k^2} \int \rd^4 x \,{\rm d}^4\q \, \cE \,  \cW^2     +{\rm c.c.}~,
\label{ChiralAction}
\eea
with $\k$ is the gravitational constant, and $\cE$ the chiral density. The integration in \eqref{ChiralAction} is carried out over the chiral subspace of the full superspace. 
Equation \eqref{PoincareEoM.b} is the equation of motion for a non-minimal Weyl multiplet comprising the Weyl multiplet coupled to the nonlinear multiplet. As shown in \cite{deWvanHvanP} at the  component level, the vector field $V_a$ in the nonlinear multiplet obeys a differential constraint. This constraint is solved by expressing the scalar $D$ of the Weyl multiplet 
in terms of the component fields of the nonlinear multiplet, which results in an unconstrained $V_a$. Using the harmonic superspace analysis of \cite{Zupnik:1998td, KTh}, it may be shown that the non-minimal Weyl multiplet is described by an unconstrained superfield prepotential $\J^\a_i$, and its conjugate, of dimension $-5/2$.  Varying the action \eqref{ChiralAction} with respect to $\J^\a_i$ is expected to give 
\bea
\d \int \rd^4 x \,{\rm d}^4\q \, \cE \,  \cW^2  \propto \int \rd^4 x \,{\rm d}^4\q  \rd^4 \bar \q\, E \,  \d \J^\a_i 
\Big(  L^{\bold a}_j \nabla_\a^j L_{\bold a}^i + \frac{3}{2} \nabla_\a^i \log \cW \Big)
\bar \cW \cW +{\rm c.c.} ~.
\label{2.57}
\eea

In absence of the nonlinear multiplet, the Weyl multiplet is described by an unconstrained real prepotential $H$ of dimension $-2$, see \cite{KTh} for the harmonic-superspace analysis.
Varying $H$ leads to 
\bea
\d \int \rd^4 x \,{\rm d}^4\q \, \cE \,  \cW^2  \propto \int \rd^4 x \,{\rm d}^4\q  \rd^4 \bar \q\, E \,  \d H \,
\bar \cW \cW~,
\eea
instead of \eqref{2.57}.
We should also remark that the explicit form of equation  \eqref{PoincareEoM.b}
is fixed by superconformal symmetry.

The equations of motion \eqref{PoincareEoM} encode highly non-trivial information regarding the supergeometry of Poincar\'e supergravity. To see this more clearly, we utilise our local superconformal freedom to adopt the gauge
\begin{align}
	\label{Poincaregauge}
	\cW = \text{const} ~, \qquad L_{\bold a}^i = \d_{\bold a}^i ~,
\end{align}
which requires one to degauge the supergeometry from conformal to Einstein superspace. Degauging the equations of motion \eqref{PoincareEoM} and making use of the gauge conditions \eqref{Poincaregauge}, we see that they imply
	\begin{align}
		\chi_\a^i &= 0~,
	\end{align}
which is the equation of motion for Poincar\'e supergravity; the background supergeometry is controlled solely by the super-Weyl tensor $W_{\a \b}$.

\section{$\cN=3$ curved supergeometries} 
\label{Section3}

In this section, we begin by reviewing the $\cN=3$ conformal superspace approach recently developed in \cite{N=3CSS}. Following this we describe how it may be degauged to yield the $\sU(3)$ \cite{Howe} and $\sSU(3)$ \cite{N=3CSS} superspaces. Due to their structural similarity with the $\cN=2$ formulations described in the previous section, here we will emphasise only their key differences.

\subsection{$\cN=3$ conformal superspace}
\label{Section3.1}

We begin with a curved superspace $\cM^{4|12}$ which is parametrised by local coordinates 
$z^{M} = (x^{m},\theta^{\m}_\imath,\bar \theta_{\dot{\mu}}^\imath)$, where $m=0, 1, 2, 3$, $\mu = 1, 2$, $\dot{\mu} = \dot{1}, \dot{2}$ and
$\imath = \underline{1}, \underline{2}, \underline{3}$. Its structure group is chosen to be $\sSU(2,2|3)$, the $\cN=3$ superconformal group.
The corresponding Lie superalgebra, $\mathfrak{su}(2,2|3)$, is spanned by the translation $P_A=(P_a, Q_\a^i ,\bar Q^\ad_i)$, Lorentz $M_{ab}$,  $R$-symmetry
$\mathbb{Y}$ and $\mathbb{J}^{i}{}_j$, dilatation $\mathbb{D}$ and the special conformal $K^A=(K^a, S^\a_i ,\bar S_\ad^i)$ generators, see appendix \ref{AppendixA} for more details. 

In order to gauge the superconformal algebra, we associate with each non-translational generator $X_{\underline A} = (M_{ab}, \mathbb{J}^i{}_j, \mathbb{Y}, \mathbb{D}, K^A)$ a connection one-form $\O^{\underline{A}} = (\O^{ab},\F^j{}_i,\F,B,\mathfrak{F}_A) = \rd z^M \O_M{}^{\underline A}$, and with $P_A$ a supervielbein one-form $E^A = \rd z^M E_M{}^A$. We assume that the supermatrix $E_M{}^A$ is non-singular, $\text{Ber}(E_M{}^A) \equiv \text{sdet}(E_M{}^A) \neq 0$, thus a unique inverse exists. The latter is denoted $E_A{}^M$ and satisfies the properties $E_A{}^M E_M{}^B = \d_A{}^B$ and $E_M{}^A E_A{}^N = \d_M{}^N$. With respect to the supervielbein basis, the connection is expressed as $\O^{\underline{A}} = E^B \O_B{}^{\underline A}$, where $\O_B{}^{\underline A} = E_B{}^M \O_M{}^{\underline A}$. The conformally covariant derivatives $\nabla_A = (\nabla_a,\nabla_\a^i,\bar{\nabla}_i^\ad)$ are then given by:\footnote{We assume that the operators $\nabla_A$ replace $P_A$ and obey the graded commutation relations, c.f. \eqref{3.2}.}
\begin{align}
	\label{6.1}
	\nabla_A = E_{A}{}^M \partial_M - \O_A{}^{\underline B} X_{\underline B}~. 
\end{align} 

By definition, the gauge group of conformal supergravity is generated by local transformations of the form
\begin{align}
	\label{66.2}
	\nabla_A' = \re^{\mathfrak{K}} \nabla_A \re^{-\mathfrak{K}} ~, \qquad
	\mathscr{K} =  \xi^B \nabla_B + \L^{\underline B} X_{\underline B} ~.
\end{align}
Additionally, given a conformally covariant tensor superfield $\mathfrak{T}$ (with its indices suppressed), its corresponding transformation law is $\mathfrak{T}' = \re^{\mathfrak{K}} \mathfrak{T}$.
Such a superfield is called primary if $K^A \mathfrak{T} =0$. Its dimension $\D$ and $\sU(1)$ charge $q$ are defined
as follows: ${\mathbb D} \mathfrak{T} = \D \mathfrak{T}$ and ${\mathbb Y} \mathfrak{T} = Q \mathfrak{T}$. 

As discussed in section \ref{Section2.1}, in order to describe conformal supergravity, certain covariant constraints must be imposed on the algebra of covariant derivatives $[\nabla_A,\nabla_B\}$. These constraints were proposed and solved in \cite{N=3CSS}, and their solution is:
\begin{subequations} \label{algebra}
	\begin{align}
		\big \{ \nabla_\a^i , \nabla_\b^j \big \} &= 2 \ve_{\a \b} \ve^{i j k} \bigg \{ \bar{W}_\ad \bar{\nabla}^\ad_k - \bar{\nabla}^\ad_k \bar{W}^{\bd} \bar{M}_{\ad \bd} + \bar{\nabla}^\ad_l \bar{W}_\ad \mathbb{J}^l{}_k - \frac{1}{4} \bar{\nabla}^\ad_k \bar{W}_\ad \Big(2\mathbb{D} + \frac13 \mathbb{Y}\Big ) \non \\
		& \qquad \qquad \qquad - \frac \ri 2 \nabla_\g{}^\ad \bar{W}_\ad S^\g_k - \frac 1 4 \bar{\nabla}^\ad_k \bar{\nabla}_l^\bd \bar{W}_\bd \bar{S}_\ad^l + \frac 1 2 \nabla^{\g \bd} \bar{\nabla}^\gd_k \bar{W}_\bd K_{\g \gd} \bigg \} ~, \\	%
		\big \{ \nabla_\a^i , \bar{\nabla}_j^\bd \big \} &= - 2 \ri \d^i_j \nabla_\a{}^\bd ~.	%
	\end{align}
\end{subequations}
This geometry is described in terms of the super-Weyl spinor $W_\a$, which is a chiral, primary superfield of dimension $1/2$ and $\sU(1)_R$ charge $-3$
\begin{align}
	\label{SWeyl}
	\bar{\nabla}^\ad_i W_\a = 0 ~, \qquad K^B W_\a = 0 ~, \qquad \mathbb{D} W_\a = \hf W_\a ~, \qquad \mathbb{Y} W_\a = -3 W_\a~.
\end{align}
It satisfies the Bianchi identity
\begin{align}
	\ri \ve_{jkl} \nabla^{ \a i} \nabla_\a^k \nabla^{\b l} W_\b - \frac{\ri}{3} \d^i_j \ve_{klm} \nabla^{ \a k} \nabla_\a^l \nabla^{\b m} W_\a  = \ri \ve^{ikl} \bar{\nabla}_{\ad j} \bar{\nabla}_k^\ad  \bar{\nabla}_{\bd l} \bar{{W}}^\bd - \frac \ri 3 \d^i_j \ve^{klm} \bar{\nabla}_{\ad k} \bar{\nabla}_l^\ad  \bar{\nabla}_{\bd m} \bar{W}^\bd  ~,
\end{align}
which is equivalent to the requirement that ${B} = ({B}^i{}_j)$, defined by
\begin{align}
	\label{superBach}
	{B}^i{}_j :=
	\ri \ve_{jkl} \nabla^{ \a i} \nabla_\a^k \nabla^{\b l} W_\b - \frac{\ri}{3} \d^i_j \ve_{klm} \nabla^{ \a k} \nabla_\a^l \nabla^{\b m} W_\b  ~,
\end{align}
is Hermitian and traceless
\begin{subequations}
	\label{2.15}
	\begin{align}
		{B}^{\dagger} = {B} ~, \qquad \text{tr} \, {B} = 0 ~,\label{2.14b}
	\end{align}
	with the superconformal properties
	\begin{align}
		K^C {B}^i{}_j = 0 ~, \qquad \mathbb{D} {B}^i{}_j = 2 {B}^i{}_j ~, \qquad \mathbb{Y} {B}^i{}_j = 0~.
	\end{align}
Additionally, $B^i{}_j$ satisfies the conservation equations
	\begin{align}
		\nabla_\a^{(i} {B}^{j)}{}_k = \frac{1}{4} \d_k^{(i} \nabla_\a^{|l|} {B}^{j)}{}_l ~, \qquad \bar{\nabla}_{(i}^\ad {B}^{j}{}_{k)} = \frac{1}{4} \d_{(i}^j \bar{\nabla}_{|l|}^{\ad} {B}^l{}_{k)}~,
	\end{align}
\end{subequations}
hence it is the $\cN=3$ supersymmetric extension of the Bach tensor. Setting $B^i{}_j =0$ gives the equation of motion for $\cN=3$ conformal supergravity.

\subsection{Degauging (i): $\sU(3)$ superspace}
In complete analogy with the $\cN=2$ construction reviewed in section \ref{subsection2.2}, it may be shown that the dilatation connection $B_A$ describes purely gauge degrees of freedom. Specifically, it transforms algebraically under local special superconformal transformations
\begin{align}
	\mathfrak{K} = \Lambda_{B} K^{B} \quad \implies \quad
	\d_{\mathfrak{K}} B_{A} = - 2 \Lambda_{A} ~.
\end{align}
Thus, we may fix the gauge $B_A=0$ at the expense of losing local special superconformal freedom. In this gauge the latter's connection is no longer needed to maintain covariance of $\nabla_A$, and should be separated
\begin{align}
	\nabla_A = \mathfrak{D}_A - \mathfrak{F}_{AB} K^B~.
\end{align}

To elucidate the structure of the operator $\mathfrak{D}_A$ and superfields $\mathfrak{F}_{AB}$ appearing in this expression, it is necessary to make use of identity \eqref{4.3}. In particular, a routine calculation leads to the following expressions for the degauged special conformal connection:
\begin{subequations} \label{connections}
	\bea
	\mathfrak{F}_\a^i{}_\b^j  
	&=&
	-\hf\ve_{\a\b}S^{ij}
	- \ve^{ijk }Y_{\a\b k}
	+ \frac 1 4 \ve_{\a \b} \ve^{ijk} \bar{\mathfrak{D}}^\ad_k \bar{W}_\ad
	~,
	\\
	\mathfrak{F}^\ad_i{}^\bd_j  
	&=&
	-\hf\ve^{\ad\bd}\bar{S}_{ij}
	+ \ve_{ijk} \bar{Y}^{\ad\bd k}
	- \frac 14 \ve^{\ad \bd} \ve_{ijk} \mathfrak{D}^{\a k} W_\a
	~,\\
	\mathfrak{F}_\a^i{}^\bd_j
	&=&
	- \mathfrak{F}^\bd_j{}_\a^i
	=
	-\d^i_jG_\a{}^\bd
	-\ri G_\a{}^\bd{}^i{}_j
	~.
	\eea
\end{subequations}
The dimension-1 superfields introduced above have the following symmetry properties:  
\bea
S^{ij}=S^{ji}~, \qquad Y_{\a\b i} =Y_{\b\a i}~, \qquad {G_{\a \ad}{}^{i}{}_i} = 0~,
\eea
and satisfy the reality conditions
\bea
\overline{S^{ij}} =  \bar{S}_{ij}~,\qquad
\overline{Y_{\a\b i}} = \bar{Y}_{\ad\bd}^i~,\qquad
\overline{G_{\b\ad}} = G_{\a\bd}~,\qquad
\overline{G_{\b\ad}{}^{i}{}_j} = - G_{\a\bd}{}^j{}_{i}
~.~~~~~~
\eea
The ${\sU}(1)_R$ charges of the complex fields are:
\bea
{\mathbb Y} S^{ij}=2S^{ij}~,\qquad
{\mathbb Y}  Y_{\a\b i}=2Y_{\a\b i}~.
\eea
Now, by employing \eqref{4.3}, we find that the algebra of spinor covariant derivatives is:
\begin{subequations} \label{U(3)algebra}
	\bea
	\{ \mathfrak{D}_\a^i , \mathfrak{D}_\b^j \}
	&=&
	2 \ve_{\a \b} \ve^{ijk} \bar{W}_\ad \bar{\mathfrak{D}}^\ad_k
	+ 4 S^{ij}  M_{\a\b} 
	- 4 \ve_{\a \b} \ve^{ijk} Y^{\g \d}_k M_{\g \d}
	+ 2 \ve_{\a \b} \ve^{ijk} \bar{\mathfrak{D}}^\ad_k \bar{W}^\bd \bar{M}_{\ad \bd}
	\non \\
	&~&
	- 4\ve_{\a \b} S^{k[i} \mathbb{J}^{j]}{}_k
	+ 8 \ve^{kl(i}{Y}_{\a\b k}  \mathbb{J}^{j)}{}_l
	+ 2 \ve_{\a \b} \ve^{ijk} \bar{\mathfrak{D}}^\ad_l \bar{W}_\ad \mathbb{J}^l{}_k
	~,
	\\
	\{ \mathfrak{D}_\a^i , \bar{\mathfrak{D}}^\bd_j \}
	&=&
	- 2 \ri \d_j^i\mathfrak{D}_\a{}^\bd
	+4\Big(
	\d^i_jG^{\g\bd}
	+\ri G^{\g\bd}{}^i{}_j
	\Big) 
	M_{\a\g} 
	+4\Big(
	\d^i_jG_{\a\gd}
	+\ri G_{\a\gd}{}^i{}_j
	\Big)  
	\bar{M}^{\bd\gd}
	\non\\
	&&
	+8 G_\a{}^\bd \mathbb{J}^i{}_j
	+4\ri\d^i_j G_\a{}^\bd{}^{k}{}_l \mathbb{J}^l{}_{k}
	-\frac{2}{3}\Big(
	\d^i_jG_\a{}^\bd
	+\ri G_\a{}^\bd{}^i{}_j
	\Big)
	\mathbb{Y} 
	~.
	\eea
	\esubeq
Additionally, solving eq. \eqref{4.3} leads to a set of differential constraints on $\mathfrak{F}_{AB}$, which we interpret as Bianchi identities. Specifically:
	\begin{subequations}\label{BI-U3}
		\bea
		\bar{\mathfrak{D}}_{\ad i} W_\a &=& 0~, \\
		\mathfrak{D}_{\a}^{(i}S^{jk)}&=&0~, \\
		\bar{\mathfrak{D}}_{\ad i}S^{jk} &=& \ri\mathfrak{D}^{\b (j}G_{\b\ad}{}^{k)}{}_i + \frac{1}{4} \d_i^{(j} \Big( 2 \bar{\mathfrak{D}}_{\ad l} S^{k)l} - \ri {\mathfrak{D}}^{\b |l|} G_{\b \ad}{}^{k)}{}_l\Big) - \d^{(i}_j \ve^{k)lm} \bar{\mathfrak{D}}_{(\ad l} \bar{\mathfrak{D}}_{\bd) m} \bar{W}^\bd ~, ~~~
		\\
		\mathfrak{D}_{(\a}^{i}Y_{\b\g)j}&=& \frac 1 3 \d^i_j \mathfrak{D}_{(\a}^{k} Y_{\b\g)k}, \\
		\mathfrak{D}_\a^k S^{ij} &=& \mathfrak{D}_\a^{(i} S^{j)k} + 2 \ve^{kl(i} \Big ( \mathfrak{D}^{\b j)} Y_{\a \b l} + 3 \ri G_{\aa}{}^{j)}{}_l \bar{W}^\ad \Big )~, \\
		\mathfrak{D}^{\b i} Y_{\a \b i} &=& \frac{3 \ri}{2} (\mathfrak{D}_\aa - 6\ri G_{\aa}) \bar{W}^\ad \\
		\mathfrak{D}_{(\a}^i G_{\b)\bd} &=& \frac{1}{4} \ve^{ijk} \bar{\mathfrak{D}}_{\bd j} Y_{\a \b k} - \frac{\ri}{16} \mathfrak{D}_{(\a}^i G_{\b) \bd}{}^i{}_j ~, \\
		\mathfrak{D}^{\a i} G_{\a \ad}&=& \frac{5\ri}{16} \mathfrak{D}^{\a j} G_{\aa}{}^i{}_j + \frac 1 8 \bar{\mathfrak{D}}_{\ad j} S^{ij} - \frac 1 8 \ve^{ijk} \bar{\mathfrak{D}}_{\ad j} \bar{\mathfrak{D}}_{\bd k} \bar{W}^\bd + \bar{Y}_{\ad \bd}^i \bar{W}^\bd - \frac 3 4 \mathfrak{D}^{\a i} W_\a \bar{W}_\ad ~,~~~ \\
		\mathfrak{D}_{(\a}^{(i}G_{\b)\bd}{}^{j)}{}_k&=&\frac{1}{4} \mathfrak{D}_{(\a}^l G_{\b) \bd}{}^{(i}{}_l \d^{j)}_k~, \\
		\mathfrak{D}_{(\a}^{[i} G_{\b) \bd}{}^{j]}{}_{k} &=& - \frac{1}{2} \mathfrak{D}_{(\a}^{l} G_{\b) \bd}{}^{[i}{}_{l} \d_k^{j]} + \d^{[i}_j \ve^{k]lm} \bar{\mathfrak{D}}_{\bd l} Y_{\a\b m}\\
		\mathfrak{D}^{\a[i}G_{\a \ad}{}^{j]}{}_k&=&-\frac{1}{2} \mathfrak{D}^{\a l} G_{\a \ad}{}^{[i}{}_l \d^{j]}_k + \frac{\ri}{4} \ve^{ijl} \Big( \bar{\mathfrak{D}}_{lk} + 2 \bar{S}_{lk} \Big ) \bar{W}_\ad~, ~~~~~ 
		\eea
	\end{subequations}
	where we have made the definitions:
	\begin{align}
		\mathfrak{D}^{ij} = \mathfrak{D}^{\a (i} \mathfrak{D}_{\a}^{j)} ~, \qquad \bar{\mathfrak{D}}_{i j} = \bar{\mathfrak{D}}_{\ad (i}\bar{\mathfrak{D}}^{\ad}_{j)}~.
	\end{align}
	
Our results above show that conformal superspace may be degauged to a supergeometry described by $\mathfrak{D}_A$ with $\sSL(2,\mathbb{C}) \times \sU(3)_R$ as its structure group. Remarkably, this is exactly the $\sU(3)$ superspace of Howe \cite{Howe}.
	
Now, we describe how the residual local dilatation and special superconformal transformations of conformal superspace manifest within this framework. It may be shown that the following structure group transformations, parametrised by a dimensionless real scalar superfield $\S$ = $\bar{\S}$, preserves the gauge $B_A = 0$
	\begin{align}
		\mathfrak{K}(\S) = \S \mathbb{D} + \hf \nabla_B \S K^{B} \quad \implies \quad B'_A = 0~.
	\end{align}
At the level of $\sU(3)$ superspace, this induces the following super-Weyl transformations
	\begin{subequations}
		\label{6.17}
		\bea
		\mathfrak{D}^{' i}_\a&=&\re^{\hf \S} \Big( \mathfrak{D}_\a^i+2\mathfrak{D}^{\b i}\S M_{\a \b} + 2 \mathfrak{D}_{\a}^j \S \mathbb{J}^{j}{}_i 
		- \frac{1}{6} \mathfrak{D}_\a^i\S {\mathbb Y} \Big)
		~,
		\label{Finite_D}
		\\
		\bar{\mathfrak{D}}_{i}^{' \ad}&=&\re^{\hf \S} \Big( \bar{\mathfrak{D}}^\ad_i+2 \bar{\mathfrak{D}}_{i}^{\bd} \S \bar{M}_{\ad \bd} - 2 \bar{\mathfrak{D}}^{\ad}_j \S \mathbb{J}^{j}{}_i 
		+ \frac{1}{6} \bar{\mathfrak{D}}_{i}^{\ad} {\mathbb Y} \Big)
		~,
		\label{Finite_DB}\\
		{\mathfrak{D}}'_\aa&=&\re^{\S} \Big( \mathfrak{D}_{\a \ad} + {\rm i} \mathfrak{D}^i_{\a} \S \bar{\mathfrak{D}}_{\ad i} + {\rm i} \bar{\mathfrak{D}}_{\ad i} \S \mathfrak{D}_{\a}^i + \Big( \mathfrak{D}^\b{}_\ad \S - \ri \mathfrak{D}^{\b i} \S \bar{\mathfrak{D}}_{\ad i} \S \Big) M_{\a \b} \non \\ && + \Big( \mathfrak{D}_{\a}{}^\bd \S + \ri \mathfrak{D}_{\a}^i \S \bar{\mathfrak{D}}^{\bd}_i \S \Big) { \bar M}_{\ad \bd} 
		- 2\ri \mathfrak{D}_\a^i \S \bar{\mathfrak{D}}_{\ad j} \S \mathbb{J}^j{}_i
		+ \frac{\ri}{6} \mathfrak{D}_{\a}^i \S \bar{\mathfrak{D}}_{\ad i} \S \mathbb{Y} 
		\Big)
		~,
		\label{Finite_DBB}
		\\
		W^{'}_\a&=& \re^{\hf \S} W_\a
		\\
		S^{' ij}&=& \re^{\S} \Big( S^{ij}
		-\hf \mathfrak{D}^{ij} \S + \mathfrak{D}^{\a (i} \S \mathfrak{D}_\a^{j)} \S \Big)
		\label{Finite_S}~,
		\\
		Y^{'}_{\a\b i}&=& \re^{\S} \Big( Y_{\a\b i}
		+ \frac 14 \ve_{ijk} \mathfrak{D}_\a^{j} \mathfrak{D}_\b^{k} \S + \hf \ve_{ijk} \mathfrak{D}_\a^{j} \S \mathfrak{D}_\b^{k} \S \Big)
		\label{Finite_Y}~,
		\\
		G'_{\a\ad}&=&
		\re^{\S} \Big( G_{\a\ad}
		-{\frac{1}{12}}[\mathfrak{D}_\a^i,\bar{\mathfrak{D}}_{\ad i}]\S
		-\frac{1}{2} \mathfrak{D}_\a^i \S \bar{\mathfrak{D}}_{\ad i} \S \Big)
		~,
		\label{Finite_G}
		\\
		G'_{\a\ad}{}^{i}{}_j&=& \re^{\S} \Big( G_{\a\ad}{}^{i}{}_j
		+{\frac\ri 4} \Big( [\mathfrak{D}_\a^{i},\bar{\mathfrak{D}}_{\ad j}] - \frac{1}{3} \d^i_j [\mathfrak{D}_\a^{k},\bar{\mathfrak{D}}_{\ad k}] \Big) \S \Big)
		~.
		\eea
	\end{subequations}
	It should be mentioned that, in the infinitesimal limit $\S^2 = 0$, the corresponding super-Weyl transformations coincide with the ones of \cite{Howe}.
	
	\subsection{Degauging (ii): $\sSU(3)$ superspace}
	\label{Section3.3}
	
	In the recent work \cite{N=3CSS}, it was shown that the torsion superfield $G_\aa{}^i{}_j$ of $\sU(3)$ superspace describes purely gauge degrees of freedom.
	Thus, by employing the super-Weyl freedom described by eq. \eqref{6.17}, it may be gauged away
	\bea
	G_{\aa}{}^{i}{}_j=0~.
	\label{3.19}
	\eea
	In this gauge, it is natural introduce new covariant derivatives $\mathscr{D}_A$
	\bea
	\label{6.19}
	\mathfrak{D}_a = \mathscr{D}_a - \frac{\ri}{3} G_a {\mathbb Y}~, \qquad \mathfrak{D}_\a^i = \mathscr{D}_\a^i~, \qquad \bar{\mathfrak{D}}_i^\ad = \bar{\mathscr{D}}_i^\ad~.
	\eea
	Then, by making use of \eqref{U(3)algebra}, we find that these covariant derivatives obey the algebra:
	\begin{subequations} 
		\label{4.21}
		\bea
		\{ \mathscr{D}_\a^i , \mathscr{D}_\b^j \}
		&=&
		2 \ve_{\a \b} \ve^{ijk} \bar{W}_\ad \bar{\mathscr{D}}^\ad_k
		+ 4 S^{ij}  M_{\a\b} 
		- 4 \ve_{\a \b} \ve^{ijk} Y^{\g \d}_k M_{\g \d}
		+ 2 \ve_{\a \b} \ve^{ijk} \bar{\mathscr{D}}^\ad_k \bar{W}^\bd \bar{M}_{\ad \bd}
		\non \\
		&~&
		- 4\ve_{\a \b} S^{k[i} \mathbb{J}^{j]}{}_k
		+ 8 \ve^{kl(i}{Y}_{\a\b k}  \mathbb{J}^{j)}{}_l
		+ 2\ve_{\a \b} \ve^{ijk} \bar{\mathscr{D}}^\ad_l \bar{W}_\ad \mathbb{J}^l{}_k~,
		\label{acr1} \\
		\{\mathscr{D}_\a^i,\cDB^\bd_j\}&=&
		-2\ri\d^i_j \mathscr{D}_\a{}^\bd
		+4\d^{i}_{j}G^{\d\bd}M_{\a\d}
		+4\d^{i}_{j}G_{\a\gd}\bar{M}^{\gd\bd}
		+8 G_\a{}^\bd \mathbb{J}^{i}{}_{j}~.~~~~~~~~~ 
		\eea
	\end{subequations}
	It should be emphasised that the Bianchi identities associated with \eqref{4.21} may be read off from \eqref{BI-U3} by keeping in mind equations \eqref{3.19} and \eqref{6.19}.
	
	Now, by examining equations \eqref{4.21}, we see that the $\sU(1)_R$ curvature has been eliminated, thus $\F_A$ is flat. As a result, an appropriate local $\sU(1)_R$ transformation may be used to fix the gauge
	\begin{align}
		\label{3.22}
		\F_A = 0~.
	\end{align}
	Consequently, the structure group reduces to $\sSL(2,\mathbb{C}) \times \sSU(3)_R$. It is because of the $\sSU(3)_R$ factor that this geometry was coined `$\sSU(3)$ superspace' in \cite{N=3CSS}.
	
	It turns out that the gauge conditions \eqref{G2} and \eqref{3.22} allow for residual super-Weyl transformations, which are described
	by a parameter $\S$ constrained by
	\be
	\Big( [\mathfrak{D}_\a^{i},\bar{\mathfrak D}_{\ad j}] - \frac 1 3 \d^i_j [\mathfrak{D}_\a^{k},\bar{\mathfrak D}_{\ad k}] \Big) \S=0~.
	\label{Ucon}
	\ee
	The general solution of this condition is
	\bea
	\S = \frac{1}{2} (\s+\bar{\s})~, \qquad \bar{\mathscr{D}}^\ad_i \s =0~,
	\qquad {\mathbb Y} \s =0~,
	\eea
	To preserve the gauge condition $\F_A=0$, every super-Weyl transformation, eq. \eqref{6.17}, must be accompanied by the following compensating $\sU(1)_R$ transformation 
	\be
	\mathfrak{D}_A \longrightarrow \re^{-\frac{1}{12} (\s - \bar{\s}) \mathbb{Y}} \mathfrak{D}_A \re^{\frac{1}{12} (\s - \bar{\s}) \mathbb{Y}}~.
	\ee
	As a result, the geometry of $\sSU(3)$ superspace is preserved by the following set of super-Weyl transformations:
	\begin{subequations}
		\label{SU(3)SW}
		\begin{align}
			\mathscr{D}_\a^{'i}&= \re^{\frac{1}{6} \s + \frac 1 3 \bar{\s}} \Big( \mathscr{D}_\a^i+ \mathscr{D}^{\b i}\s M_{\a \b} - \mathscr{D}_{\a}^j\s \mathbb{J}^{i}{}_j \Big) ~, 
			\\ 
			\bar{\mathscr{D}}_{i}^{' \ad}&=\re^{\frac{1}{3} \s + \frac{1}{6} \bar{\s}} \Big( \bar{\mathscr{D}}^\ad_i-2 \bar{\mathscr{D}}_{ \bd i} \bar{\s} \bar{M}^{\ad \bd} + \bar{\mathscr{D}}^{\ad}_j \bar{\s} \mathbb{J}^{j}{}_i \Big)~,
			\\
			\mathscr{D}_\aa' &= \re^{\hf \s + \hf \bar{\s}} \Big(\mathscr{D}_\aa + \frac{\rm i}{2} \mathscr{D}^i_{\a} \s \bar{\mathscr{D}}_{\ad i} + \frac{\rm i}{2} \bar{\mathscr{D}}_{\ad i} \bar{\s} \mathscr{D}_{\a}^i + \hf \Big( \mathscr{D}^\b{}_\ad (\s + \bar \s ) - \frac{\ri}{2} \mathscr{D}^{\b i} \s \bar{\mathscr{D}}_{\ad i} \bar{\s} \Big) M_{\a \b} \non \\ & + \hf \Big( \mathscr{D}_{\a}{}^\bd (\s + \bar{\s}) + \frac{\ri}{2} \mathscr{D}_{\a}^i \s \bar{\mathscr{D}}^{\bd}_i \bar{\s} \Big) { \bar M}_{\ad \bd} 
			- \frac \ri 2 \mathscr{D}_\a^i \s \bar{\mathscr{D}}_{\ad j} \bar{\s} \mathbb{J}^j{}_i \Big) ~, \\
			W'_\a&= \re^{\hf \s} W_\a
			\\
			S'^{ ij}&= \re^{\frac{1}{3} \s + \frac 2 3 \bar{\s}} \Big( S^{ij}-{\frac14}\mathscr{D}^{ij} \s + \frac 1 4 \mathscr{D}^{\a (i} \s \mathscr{D}_\a^{j)} \s \Big)~, 
			\label{super-Weyl-S} \\
			Y'_{\a\b i}&= \re^{\frac{1}{3} \s + \frac 2 3 \bar{\s}} \Big( Y_{\a\b i}+{\frac18} \ve_{ijk} \mathscr{D}_{(\a}^{j} \mathscr{D}_{\b)}^{k}\s + \frac18 \ve_{ijk} \mathscr{D}_{(\a}^{j} \s \mathscr{D}_{\b)}^{k} \s \Big)~,
			\label{super-Weyl-Y} \\
			G_{\aa}' &=
			\re^{\frac12 \s + \frac12 \bar{\s}} \Big(
			G_{\a\bd} -{\frac{\ri}4}
			\mathscr{D}_{\a \ad} (\s-\bar{\s})
			-\frac{1}{8} \mathscr{D}_\a^i \s \bar{\mathscr{D}}_{\ad i} \bar{\s}
			\Big)
			~.
			\label{super-Weyl-G}
		\end{align}
	\end{subequations}

\section{The $\cN=3$ nonlinear multiplet and supergravity}
\label{Section4}

In section \ref{N=2ESS}, we saw that it was possible to further degauge $\sSU(2)$ superspace by coupling to a nonlinear multiplet, which can then be utilised to enforce a gauge on $\sSU(2)_R$ such that the structure group reduces to $\sSL(2,\mathbb{C})$. It is our goal in this section to show that a similar construction is valid at the $\cN=3$ level. 

A key ingredient in this construction is an $\cN=3$ nonlinear multiplet, which, to the best of our knowledge, has not appeared in the literature. We propose it to be a primary, dimensionless, and uncharged matrix superfield $L^i_{\bold a}$, where $\bold{a}$ corresponds to an external $\sSU(3)$ group. Within conformal superspace, it is characterised by the superconformal constraints
\begin{subequations}
	\label{SU(3)NL}
\bea
		  L^i_{\bold{a}} \bar{L}^{\bold{b}}_i &=& \d_{\bold a}^{\bold b}  \quad \Longleftrightarrow  \quad L^i_{\bold{a}} \bar{L}^{\bold{a}}_j = \d_{i}^{j} ~,\qquad \text{det}\,L = 1 ~,  \label{SU(3)NL-a} \\
		\bar L^{\bold a}_k \nabla^{(i}_\a L^{j)}_{\bold a} &-& \frac 14  \d^{(i}_k\bar L^{\bold a}_l     \nabla^{|l|}_\a L^{j)}_{\bold a}=0~,
		\qquad 
	 \bar{L}^{\bold a}_k 	\nabla_\a^{[i} L_{\bold a}^{j]}
		- \frac 12 \d_k^{[i}  \bar{L}^{\bold a}_l     \nabla_\a^{|l|} L_{\bold a}^{j]} =0~,
		\label{NLDiff}
\eea
\end{subequations}
where $\bar{L}^{\bold a}_i = \overline{L_{\bold a}^i}$. 
The constraints \eqref{SU(3)NL-a} mean that $L := ( L^i_{\bold{a}} ) \in \sSU(3)$, and thus
\bea
\bar{L}^{\bold{a}}_i = \hf \ve^{\bold{a} \bold{b} \bold{c}} \ve_{ijk} L^j_{\bold{b}} L^k_{\bold{c}}~.  
\eea
The left-hand sides of the constraints \eqref{NLDiff} are primary superfields. These constraints can be recast in the form
\bea
\bar L^{\bold a}_k \nabla^{i}_\a L^{j}_{\bold a} 
= \frac 14 \d^{(i}_k  \bar L^{\bold a}_l    \nabla^{|l|}_\a L^{j)}_{\bold a}
+  \frac 12  \d_k^{[i}  \bar{L}^{\bold a}_l    \nabla_\a^{|l|} L_{\bold a}^{j]} ~.
\eea

\subsection{$\cN=3$ Einstein superspace}

Within the $\sSU(3)$ superspace, which will be the starting point of our analysis, the nonlinear multiplet is subject to the constraints \eqref{SU(3)NL} with the replacement $\nabla \rightarrow \mathscr{D}$. We begin by fixing the local $\sSU(3)_R$ freedom by adopting the gauge
\begin{align}
	\label{SU(3)gauge}
	L_{\bold a}^i = \d_{\bold{a}}^i~.
\end{align}
Since this constraint fixes the $\sSU(3)_R$ freedom, the corresponding connection in $\mathcal{D}_A$ should be separated
\begin{align}
	\mathscr{D}_A = \cD_A + \Phi_A{}^{j}{}_k \mathbb{J}^{k}{}_j ~, \qquad \cD_A = E_A{}^M \partial_M + \hf \O_A{}^{bc} M_{bc}~.
\end{align}
Consistency with the differential constraints \eqref{NLDiff} leads to the following conditions on the $\sSU(3)_R$ connection
\begin{align}
	\Phi_\a^{(i j)}{}_k - \frac 1 4 \d_k^{(i} \F_{\a,}^{|l| , j)}{}_l = 0 ~, \qquad \F_{\a }^{[i ,j]}{}_k - \frac 1 2 \d_k^{[i} \F_{\a}^{|l| , j]}{}_l = 0~.
\end{align}
These imply that the degauged spinor covariant derivative $\cD_\a^i$ takes the form\footnote{It may also be shown that $\mathscr{D}_a = \cD_a$ in the gauge \eqref{SU(3)gauge}.}
\begin{align}
	\label{SU(3)degauged}
	\mathscr{D}_\a^i = \cD_\a^i + \frac 3 4 \chi_\a^j \mathbb{J}^{i}{}_j
\end{align}
where $\chi_\a^i$ is a dimension-$1/2$ superfield. This allows us to deduce the following relations:
\begin{subequations}
	\begin{align}
		\{ \cD_\a^i , \cD_\b^j \} &= \{ \mathscr{D}_\a^i , \mathscr{D}_\b^j\} + \chi_{(\a}^{(i} \cD_{\b)}^{j)} - \ve_{\a\b} \chi^{\g [i} \cD_\g^{j]} - \frac 3 2 \cD_{(\a}^{(i} \chi_{\b)}^{|k|} \mathbb{J}^{j)}{}_k \non \\
		&\phantom{=} - \frac 3 4 \ve_{\a \b} (\cD^{\g [i} \chi_\g^{|k|}) \mathbb{J}^{j]}{}_k - \frac 3 8 \chi_{(\a}^{(i} \chi_{\b)}^{|k|} \mathbb{J}^{j)}{}_k - \frac 3 {16} \ve_{\a \b} \chi^{\g [i} \chi_\g^{|k|} \mathbb{J}^{j]}{}_{k}~, \\
		\{ \cD_{\a}^{i} , \cDB^{\ad}_{j} \} &= \{ \mathscr{D}_{\a}^{i} , \bar{\mathscr{D}}^{\ad}_{j} \} + \frac 1 4 \Big(\bar{\chi}^\ad_j \cD_\a^i - 3 \d_j^i \bar{\chi}^\ad_k \cD_\a^k \Big) + \frac 1 4 \Big ( \chi_\a^i \cDB^\ad_j - 3 \d^i_j \bar{\chi}^\ad_k \cD_\a^k \Big ) + \frac 3 4 (\cD_\a^i \bar{\chi}^\ad_k) \mathbb{J}^k{}_j \non \\
		&\phantom{=} - \frac 3 4 (\cDB^\ad_j \chi_\a^k) \mathbb{J}^i{}_k - \frac 3 {16} \chi^i_\a \bar{\chi}^\ad_k \mathbb{J}^k{}_j - \frac 3 {16} \chi_\a^k \bar{\chi}^\ad_j \mathbb{J}^i{}_k + \frac 9 {16} \chi_\a^k \bar{\chi}^\ad_k \mathbb{J}^i{}_j ~.
	\end{align}
\end{subequations}

Since the graded commutators $[\cD_A, \cD_B\}$ do not contain $\sSU(3)_R$ curvature, it is clear that such contributions above identically vanish, which has highly nontrivial implications. In particular, we find that the torsions $S^{ij}$, $Y_{\a \b i}$ and $G_\aa$ of $\sSU(3)$ superspace become descendants of the spinor torsion $\chi_\a^i$:
\begin{subequations}
	\begin{align}
		S^{ij} &= - \frac 3 {16} \cD^{\a (i} \chi_\a^{j)} - \frac 3 {64} \chi^{\a(i} \chi_\a^{j)}~, \\
		Y_{\a \b i} &= \frac 3 {32} \ve_{ijk} \Big( \cD_{(\a}^j \chi_{\b)}^k + \frac 1 4 \chi_{(\a}^j \chi_{\b)}^k \Big)~, \\
		G_{\aa} &=  \frac 1 {32} \Big( \cDB_{\ad i} \chi_\a^i - \cD_\a^i \bar{\chi}_{\ad i} - \frac{7}{4} \chi_\a^i \bar{\chi}_{\ad i} \Big)~.
	\end{align}
\end{subequations} 
We also find that $\chi_\a^i$ and $W_\a$ are constrained by the Bianchi identities
\begin{subequations}
\begin{align}
	\cDB^\ad_i W_\a &= 0 ~, \qquad \cD^{\a i} {W}_\a - \frac 3 4 {\chi}^{\a i} {W}_\a = 0~, \\
	\cD_{(\a}^{(i} \chi_{\b)}^{j)} &= 0 ~, \qquad \ve_{ijk} \cD^{\a j} \chi_\a^k = 0~, \\
	\cD_\a^{i} \bar{\chi}_{\ad j} - \frac 1 4 &\chi_\a^{i} \bar{\chi}_{\ad j} - \frac{1}{3} \d^i_j \Big(\cD_\a^k \bar{\chi}_{\ad k} - \frac 1 4 \chi_\a^{k} \bar{\chi}_{\ad k}\Big) = 0~.
\end{align}
\end{subequations}
As a result, we find the algebra of degauged covariant derivatives $\cD_A$ to be:
\begin{subequations} \label{N=3EinsteinGeometry}
	\begin{align}
		\{ \cD_\a^i , \cD_\b^j \} &= \chi_{(\a}^{(i} \cD_{\b)}^{j)} - \ve_{\a\b} \chi^{\g [i} \cD_\g^{j]} + 2 \ve_{\a \b} \ve^{ijk} \bar{W}_\ad \cDB^\ad_k + 4 S^{ij} M_{\a \b} \non \\
		&\phantom{=} + 2 \ve_{\a \b} \ve^{i jk} \Big ( 4 Y^{\g \d}_k M_{\g \d} + \cDB^{\gd}_k\bar{W}^{\d} \bar{M}_{\gd \dd} \Big ) ~, \\
		\{ \cD_\a^i , \cDB^\ad_j \} &=  \frac 1 4 \Big(\bar{\chi}^\ad_j \cD_\a^i - 3 \d_j^i \bar{\chi}^\ad_k \cD_\a^k \Big) + \frac 1 4 \Big ( \chi_\a^i \cDB^\ad_j - 3 \d^i_j \bar{\chi}^\ad_k \cD_\a^k \Big ) \non \\
		&\phantom{=} - 2 \ri \d^i_j \cD_\a{}^{\ad} + 4 \d^i_j \Big( G^{\b \ad} M_{\a \b} + G_{\a \bd} \bar{M}^{\ad \bd} \Big)~.
	\end{align}
\end{subequations}
We emphasise that this geometry is described solely in terms of the spinors $\chi_\a^i$ and $W_{\a}$. Additionally, it should be pointed out that $\mathscr{D}_{\aa} = \cD_{\aa}$.

As was the case for the $\cN=2$ Einstein superspace studied in section \ref{N=2ESS}, the supergravity gauge freedom of this geometry includes both local $\cK$ and super-Weyl transformations. The former act on the covariant derivatives $\cD_A$ and tensor superfields $\cT$ as follows
\begin{subequations}
\end{subequations}
\begin{align}
	\d_\cK \cD_A &= [\cK, \cD_A] ~, \qquad \d_\cK = \cK \cT \non \\
	\cK &= \xi^B \cD_B + \frac 1 2 K^{ab} M_{ab}~.
\end{align}
Additionally, keeping in mind equations \eqref{SU(3)SW} and \eqref{SU(3)degauged}, it follows that the super-Weyl transformations of this supergeometry take the form
\bsubeq
\bea
{\cD'}_\a^i&=& \re^{\frac 1 6 \s + \frac 1 3 \bar{\s} } \Big( \cD_\a^i+\cD^{\b i}\s M_{\b \a} \Big)~, 
\\
{\bar{\cD}'}_{\ad i}&=&\re^{\frac 1 6 \s + \frac 1 3 \bar{\s} } \Big( \bar{\cD}_{\ad i}+\bar{\cD}^{\bd}_{i}\sba \bar{M}_{\bd\ad} \Big)~, \\
{\cD'}_{ \! \! \aa}&=& \re^{\hf \s + \hf \bar{\s}} \Big(\cD_\aa + \frac{\rm i}{2} \cD^i_{\a} \s \cDB_{\ad i} + \frac{\rm i}{2} \cDB_{\ad i} \bar{\s} \cD_{\a}^i + \hf \Big( \cD^\b{}_\ad (\s + \bar \s ) - \frac{\ri}{2} \cD^{\b i} \s \cDB_{\ad i} \bar{\s} \Big) M_{\a \b} \non \\ 
&& \qquad \qquad + \hf \Big( \cD_{\a}{}^\bd (\s + \bar{\s}) + \frac{\ri}{2} \cD_{\a}^i \s \cDB^{\bd}_i \bar{\s} \Big) { \bar M}_{\ad \bd} \Big) ~, \\
{\chi'}_\a^i &=& \re^{\frac 1 6 \s + \frac 1 3 \bar{\s} } \Big ( \chi_\a^i + \frac 4 3\cD_\a^i \s \Big )~, \\
{W}'_{\a}&=& \re^{\frac 1 2 \s} {W}_{\a}~,
\eea
\end{subequations}
where $\s$ is a chiral parameter, $\bar{\cD}_i^\ad \s = 0$.

A comment is in order. Had we undertaken the analysis of this subsection by beginning from $\sU(3)$ superspace instead of $\sSU(3)$ superspace, the resulting algebra of covariant derivatives would still be expressed solely in terms of $\chi_\a^i$ and $W_\a$. Specifically, in this situation the superfield $G_{\aa}{}^i{}_j$ is no longer independent; it will be a descendant of $\chi_\a^i$.

On another note, there is an additional point regarding the nonlinear multiplet which should be emphasised. Specifically, one may omit the second constraint in eq. \eqref{NLDiff}, resulting in a larger multiplet. Repeating the analysis above with this multiplet then leads to a supergeometry described by a primary dimension-1/2 superfield $\vf_{\a ij} = \vf_{\a ji}$, in addition to $W_\a$ and $\chi_\a^i$. It is unclear to us what the role of $\vf_{\a ij}$ is.

\subsection{$\cN=3$ Poincar\'e supergravity}

Similar to the $\cN=2$ case described in section \ref{section2.5}, we will show how the geometric setup developed in this section can be utilised to provide a powerful description of Poincar\'e supergravity. To this end, we begin by coupling a background conformal superspace to a vector multiplet $\cW^i$ and nonlinear multiplet $L_{\bold a}^i$. The former is necessarily on-shell by virtue of its defining constraints
\begin{align}
	\nabla_\a^{(i} \cW^{j)} = 0~, \qquad \bar{\nabla}_{i}^\ad \cW^j = \frac 1 3 \d_i^j \bar{\nabla}^\ad_k \cW^k ~,
\end{align}
while the latter solves eq. \eqref{SU(2)NL}. On-shell the nonlinear multiplet satisfies the superconformal constraint
\begin{align}
	\label{PoincareEoMN=3}
	& \bar{L}^{\bold a}_j \nabla_\a^j L_{\bold a}^i + \frac{8}{9} \nabla_\a^i \log (|\cW|^2) = 0 ~, \qquad 
	|\cW|^2 = \cW^i \bar{\cW}_i ~.
\end{align}

The equations of motion given above encode highly non-trivial information regarding the supergeometry of Poincar\'e supergravity. To see this more clearly, we utilise our local superconformal freedom to adopt the gauge\footnote{Unlike the $\cN=2$ case, our gauge choice does not fix the $\sU(1)_R$ freedom since $|\cW|$ is real. However, it may be shown that the equation of motion \eqref{PoincareEoMN=3} implies that the $\sU(1)_R$ connection is flat.}
\begin{align}
	\label{PoincaregaugeN=2}
	|\cW| = 1 ~, \qquad L_{\bold a}^i = \d_{\bold a}^i ~,
\end{align}
which requires one to degauge the supergeometry from conformal to Einstein superspace. Performing a routine degauging procedure and making use of the gauge conditions \eqref{Poincaregauge}, we see that the equation of motion for Poincar\'e supergravity is
\begin{align}
	\chi_\a^i &= 0~,
\end{align}
hence the background supergeometry is dictated by the super-Weyl spinor $W_{\a}$. Additionally, we note that the constraint $|\cW|= 1$ implies that $\cW^i$ constitute coordinates on $S^3$.

It is interesting to see how the on-shell condition \eqref{PoincareEoMN=3} implies the equation of motion for the nonlinear multiplet. For simplicity we will work in conformally flat backgrounds, $W_\a=0$. By making use of the constraints \eqref{SU(3)NL}, one may derive the constraint
\begin{align}
	\label{4.19}
	\nabla^{\aa} \nabla_{\aa} L_{\bold a}^i &= - \frac{49}{256} \bigg\{ \nabla^{\a i} \bar{\nabla}^{\ad}_j \nabla_\a^j \bar{\nabla}_{\ad k} L_{\bold a}^k - \frac 1 3 \nabla^{\a i} \bar{\nabla}^{\ad}_j \Big( L_{\bold a}^j \bar{L}^{\bold b}_k \nabla_\a^k \bar{\nabla}_{\ad m} L_{\bold b}^m \Big) \non \\
	&\quad + \frac{16 \ri}{21} \nabla^{\aa} \Big( L_{\bold a}^i \bar{L}^{\bold b}_j \nabla_\a^k \bar{\nabla}_{\ad l} L_{\bold b}^l + L_{\bold a}^j \cX_{\aa j}{}^i \Big) - \nabla^{\a i} \bar{\nabla}^{\ad}_j (L_{\bold a}^k \cX_{\aa k}{}^j )\bigg \}~,
\end{align}
where we have introduced the descendant
\begin{align}
	\cX_{\aa j}{}^i &= \frac{1}{63}  \bigg\{ \bar{L}^{\bold a}_l \bar{L}^{\bold b}_k \nabla_\a^k L_{\bold b}^i \bar{\nabla}_{\ad j} L_{\bold a}^l + \nabla_\a^i \bar{L}_l^{\bold a} \bar{\nabla}_{\ad j} L_{\bold a}^l - 24 \bar{L}^{\bold a}_j \bar{L}^{\bold b}_l \bar{\nabla}_{\ad k} L_{\bold a}^i \nabla_\a^l L_{\bold b}^k \non \\
	&\qquad + 24 \bar{L}^{\bold a}_j \bar{L}^{\bold b}_l \bar{\nabla}_{\ad k} L_{\bold a}^k \nabla_\a^l L_{\bold b}^i - 24 \d^{i}_j \bar{\nabla}_{\ad k} \bar{L}^{\bold a}_l \nabla_\a^l L_{\bold a}^k - 24 \bar{\nabla}_{\ad j} \bar{L}^{\bold b}_k \nabla_\a^k L_{\bold b}^i \bigg \}~.
\end{align}
Now, fixing the super-Weyl gauge $|\cW| = 1$ and making use of the on-shell condition \eqref{PoincareEoMN=3}, it may be shown that eq. \eqref{4.19} reduces to
\begin{align}
	\nabla^{a} \nabla_{a} L_{\bold a}^i = 0~.
\end{align}

\section{Discussion and generalisations}
\label{Section5}

As discussed in section \ref{Section2}, the constraints \eqref{SU(2)NL} defining the $\cN=2$ nonlinear multiplet imply at the component level that the real vector field $V^a$ in the nonlinear multiplet obeys a differential nonlinear constraint \cite{deWvanHvanP} 
\bea
\nabla_a V^a - \frac 14 B| - \hf V^a V_a + \dots =0~, 
\label{ConstrainedV} 
\eea
where the ellipsis stands for the terms involving the scalar and spinor fields of the nonlinear multiplet, 
and $B|$ denotes the $\q$-independent component of the super-Bach scalar \eqref{N=2Bach}.
The auxiliary field $D$ in the Weyl multiplet \cite{deWvanHvanP}  is related to $B|$ as $B|= 12 D$. The differential equation \eqref{ConstrainedV} was interpreted in \cite{deWvanHvanP}  as an algebraic equation that expressed $D$ in terms of the component fields of the nonlinear multiplet. As a result, 
they ended up with a non-minimal $32+32$ Weyl multiplet comprising the standard Weyl multiplet coupled to the nonlinear multiplet. 
This idea has been applied, over many years, to construct dilaton Weyl multiplets in six \cite{BSVanP, Hutomo:2022hdi}, five  \cite{Bergshoeff1,  Hutomo:2022hdi} and four 
\cite{Butter:2017pbp, Gold:2022bdk} dimensions. Instead of the {\it off-shell} nonlinear multiplet, these works made use of {\it on-shell} compensating multiplets, specifically vector and scalar ones. 

Let us briefly discuss variant $\cN =2$ Weyl multiplets in four dimensions \cite{Butter:2017pbp, Gold:2022bdk}.
The dilaton Weyl multiplet proposed in  \cite{Butter:2017pbp} makes use of a vector multiplet obeying the equation of motion \eqref{PoincareEoM.a}. At the component level, the differential equations on the  fields of the vector multiplet can be interpreted as algebraic equations on certain covariant component fields of the standard Weyl multiplet, and thus the latter fields become composite.\footnote{In particular, the equation of motion \eqref{PoincareEoM.a} has the following implication 
$4 \nabla^a \nabla_a \cW  +W^{\a \b} \nabla_{\a \b} \cW + B \cW +2 \nabla^i_\a W^{\a\b} \nabla_{\b i} \cW=0$ derived in \cite{Butter:2012xg}. This equation allows $B$ to be expressed in terms of $\cW$ and its covariant derivatives.}
 This leads to a new $24+24$ Weyl multiplet of conformal supergravity.  

The dilaton Weyl multiplet proposed in \cite{Gold:2022bdk} makes use of a hypermultiplet $\U^i$ obeying the equation of motion 
\bea
\label{FSHM1}
\nabla_{\a}^{(i} \U^{j)} = 0~, \qquad \bar \nabla_{\ad}^{(i} \U^{j)} = 0~.
\eea
Again, at the component level, the differential equations on the fields of the hypermultiplet can be interpreted as algebraic equations on certain covariant component fields of the standard Weyl multiplet, and thus the latter fields become composite. For example, \eqref{FSHM1}
implies the following relation
\bea
\big( \nabla^a \nabla_a +\frac 18 B\big)\U^i = 0~, 
\label{FSHM2}
\eea
and therefore the super-Bach scalar becomes a composite superfield, 
\bea
B = - 8\frac{\bar \U_i  \nabla^a \nabla_a\U^i }{  |\U|^2} ~, \qquad |\U|^2 = \bar \U_i \U^i~.
\eea
 This results in a novel  $24+24$ Weyl multiplet of conformal supergravity.  

We believe that our $\cN=3$ nonlinear multiplet defined by the superconformal constraints 
\eqref{SU(3)NL} can be used to generate a variant $\cN=3$ Weyl multiplet. This will be discussed elsewhere. It should be pointed out that on-shell $\cN=3$ vector multiplets have recently been used to construct a dilaton Weyl multiplet for conformal supergravity and higher-derivative couplings \cite{Hegde:2022wnb, Adhikari:2024qxg}. 

A remarkable feature of the $\cN=3$ Einstein superspace geometry \eqref{N=3EinsteinGeometry} is the fact that $\cN=3$ Minkowski superspace is its unique maximally supersymmetric background. 
This follows from the theorem proved in \cite{Kuzenko:2014eqa}. The same theorem implies that the only maximally supersymmetric backgrounds for the $\cN=2$ Einstein superspace geometry \eqref{N=2EinsteinGeometry} are described by the conditions 
\bea
\c_\a^i =0~, \qquad \cD_A W_{\b\g}=0~.
\eea
It follows that no maximally supersymmetric AdS backgrounds are allowed.
It should be pointed out that a comprehensive analysis of maximally supersymmetric backgrounds in 
$\cN=2$ supergravity was given in \cite{Butter:2015tra}.

As an application of the geometric approaches to $\cN=2$ (conformal) supergravity reviewed in section \ref{Section2}, we will now show the emergence of a cosmological constant in $\cN=2$ supergravity from a superspace approach. We recall that the construction of $\cN=2$ supergravity with a cosmological term was first achieved at the component level within the superconformal tensor calculus in \cite{deWvanHvanP}. Key to their analysis was the introduction of an $\sSO(2)$ vector multiplet, which, along with the nonlinear multiplet, played the role of a conformal compensator.

Following the general philosophy of \cite{deWvanHvanP}, our first step is to couple the conformal superspace background reviewed in section \ref{Section2.1} to an $\sSO(2)$ vector multiplet. The corresponding local $\sSO(2)$ transformations act only on the `second isospinor' index $\bold a $ of the nonlinear multiplet $L_{\bold{a}}^i$. This means that we introduce new operators $\bm{\nabla}_A$ defined by
\begin{align}
	\bm{\nabla}_A = \nabla_A + \ri V_A \mathbb{Z}~,
\end{align}
where the $\sSO(2)$ generator $\mathbb{Z}$ acts on $L_{\bold{a}}^i$ in accordance with the rule
\begin{align}
	\mathbb{Z} L_{\bold a}^i = \frac{\ri}{2} (\s^2)_{\bold a}{}^{\bold b} L_{\bold b}^i~, 
	\qquad \s^2 =\left(\begin{array}{c c}
		0 & - \ri \\
		\ri & 0 \\
	\end{array}\right) ~.
\end{align}
The algebra of charged covariant derivatives $\bm{\nabla}_A$ is defined by the relations
\begin{subequations}
	\begin{align}
		\{ \bm{\nabla}_\a^i , \bm{\nabla}_\b^j \} &= 2 \ve^{ij} \ve_{\a\b} \big( \bar{W}_{\gd\dd} \bar{M}^{\gd\dd} + \frac 1 4 \bar{\bm{\nabla}}_{\gd k} \bar{W}^{\gd\dd} \bar{S}^k_\dd - \frac 1 4 \bm{\nabla}_{\g\dd} \bar{W}^\dd{}_\gd K^{\g \gd} - \ri \bar{\cW} \mathbb{Z} \big)~, \\
		\{ \bm{\nabla}_\a^i , \bar{\bm{\nabla}}^\bd_j \} &= - 2 \ri \d_j^i \bm{\nabla}_\a{}^\bd~,
	\end{align}
\end{subequations}
where $\cW$ is the field strength of the vector multiplet and is subject to the constraints
\begin{align}
	\bar{\Nabla}^\ad_i \cW = 0 ~, \qquad \Nabla^{ij} \cW = \bNabla^{ij} \bar{\cW}~.
\end{align}

To describe supergravity, it is necessary to couple the background to two compensators, which are necessary to break the dilatation, special superconformal, and $\sU(2)_R$ symmetries. Below, this role will be played by the vector multiplet $\cW$ and nonlinear multiplet $L_{\bold a}^i$. We emphasise that, in this supergeometry, $L_{\bold a}^i$ is characterised by the nonlinear conditions
\begin{align}
	\text{det}\, L = 1 ~, \qquad L^{\bold a (i} \Nabla^{j}_\a L^{k)}_{\bold a} = 0~.
\end{align}
On-shell, these compensators satisfy the primary constraints
\begin{subequations}
	\label{SGEoM}
	\begin{align}
		& L^{\bold a}_j \Nabla_\a^j L_{\bold a}^i + \frac{3}{2} \Nabla_\a^i \log \cW = 0~, \label{4.6a} \\
		& \Big ( \Nabla^{ij} - \frac{2}{3} \Big( L^{\bold a}_k \Nabla^{k(i} L_{\bold a}^{j)} + L^{\bold a}_k \Nabla^{\a k}  L_{\bold a}^{(i} \Nabla_{\a}^{j)} 
		+ \frac 1 2 L^{\bold{a} (i} \Nabla^{\a}_k L_{\bold a}^{j)} \Nabla_\a^k \Big) \Big) \cW = 0~, \label{4.6b}
 	\end{align}
\end{subequations}
whose form we deduce via symmetry principles.\footnote{Specifically, we computed the most general ans\"atze for the equations of motion \eqref{SGEoM} and required them to be superconformal. This uniquely fixed eq. \eqref{4.6a}, while eq. \eqref{4.6b} was determined up to a parameter. This was then fixed by requiring consistency with the off-shell results of appendix \ref{AppendixB}.}

The equations of motion \eqref{SGEoM} encode highly non-trivial information regarding the background supergeometry. To see this overtly, we utilise our local superconformal and $\sSO(2)$ freedom to adopt the gauge
\begin{align}
	\label{gauge}
	\cW = \text{const} ~, \qquad L_{\bold a}^i = \d_{\bold a}^i ~,
\end{align}
which requires one to degauge the supergeometry from conformal to Einstein superspace, see section \ref{Section2} for details. Degauging the equations of motion \eqref{SGEoM} and making use of the gauge conditions \eqref{gauge}, we see that they imply
\begin{subequations}
\begin{align}
	\chi_\a^i &= 0~, \label{4.8a} \\
	S^{ij} &= \frac{1}{6} \d^{ij} \cW - \frac{1}{12} \d^{k(i} {\cD}_k^\a V_\a^{j)} - \frac{1}{24} V^{\a i} V_{\a}^j - \frac{1}{72} \d^{ik} \d^{jl} V^{\a}_k V_{\a l}~. \label{4.8b}
\end{align}
\end{subequations}
Importantly, the latter equation indicates that $S^{ij}$ contains a non-zero constant term, which implies that this supergeometry is characterised by a non-zero cosmological constant, in agreement with the results of \cite{deWvanHvanP}. It should be noted, however, that a direct analysis of the supergeometry is required to confirm this result. Such an analysis proves to be quite non-trivial, and is provided in appendix \ref{AppendixB}.

To conclude, we will sketch an $\cN=3$ generalisation of this construction, which is expected to describe $\cN=3$ supergravity with a cosmological constant. To this end, we introduce the gauge-covariant derivatives $\Nabla_A$ defined by
\begin{align}
	\Nabla_A = \nabla_A + \ri V_A \mathbb{Z}~.
\end{align}
Here $\mathbb{Z}$ is an $\sSO(2) \subset \sSU(3)$ generator acting on the `second isospinor' index $\bold{a}$ of the nonlinear multiplet $L_{\bold a}^i$, which, in this supergeometry, is characterised by the constraints \begin{subequations}
	\bea
	L^i_{\bold{a}} \bar{L}^{\bold{b}}_i &=& \d_{\bold a}^{\bold b}  \quad \Longleftrightarrow  \quad L^i_{\bold{a}} \bar{L}^{\bold{a}}_j = \d_{i}^{j} ~,\qquad \text{det}\,L = 1 ~, \\
	\bar L^{\bold a}_k \Nabla^{(i}_\a L^{j)}_{\bold a} &-& \frac 14  \d^{(i}_k\bar L^{\bold a}_l     \Nabla^{|l|}_\a L^{j)}_{\bold a}=0~,
	\qquad 
	\bar{L}^{\bold a}_k 	\Nabla_\a^{[i} L_{\bold a}^{j]}
	- \frac 12 \d_k^{[i}  \bar{L}^{\bold a}_l     \Nabla_\a^{|l|} L_{\bold a}^{j]} =0~.
	\eea
\end{subequations}
The algebra of charged covariant derivatives $\bm{\nabla}_A$ is defined by the relations
\begin{subequations}
	\begin{align}
		\{ \bm{\nabla}_\a^i , \bm{\nabla}_\b^j \} &= 2 \ve_{\a \b} \ve^{i j k} \bigg \{ \bar{W}_\ad \bar{\bm{\nabla}}^\ad_k - \bar{\bm{\nabla}}^\ad_k \bar{W}^{\bd} \bar{M}_{\ad \bd} + \bar{\bm{\nabla}}^\ad_l \bar{W}_\ad \mathbb{J}^l{}_k - \frac{1}{4} \bar{\bm{\nabla}}^\ad_k \bar{W}_\ad \Big(2\mathbb{D} + \frac13 \mathbb{Y}\Big ) \non \\
		& \qquad \qquad  - \frac \ri 2 \bm{\nabla}_\g{}^\ad \bar{W}_\ad S^\g_k - \frac 1 4 \bar{\bm{\nabla}}^\ad_k \bar{\bm{\nabla}}_l^\bd \bar{W}_\bd \bar{S}_\ad^l + \frac 1 2 \bm{\nabla}^{\g \bd} \bar{\bm{\nabla}}^\gd_k \bar{W}_\bd K_{\g \gd} - \ri \bar{\cW}_k{} \mathbb{Z} \bigg \}  ~,  \\
		\{ \bm{\nabla}_\a^i , \bar{\bm{\nabla}}^\bd_j \} &= - 2 \ri \d_j^i \bm{\nabla}_\a{}^\bd~,
	\end{align}
\end{subequations}
where $\cW^{i}$ is the field strength of the vector multiplet. It satisfies the on-shell constraints
\begin{align}
	\label{VMcomp}
	\Nabla^{(i}_\a \cW^{j) } = 0 ~, \qquad \bar{\Nabla}^\ad_i \cW^{j } = \frac{1}{3} \d^j_i \bar{\Nabla}^\ad_k \cW^{k} ~.                                                                                                                                                                                                                                                                                                                                                                                                                                                                                                                                                                                                                                                                                                                                                                                                                                                                                                                                                                                                                                                   
\end{align}
We also point out that the on-shell constraints for the nonlinear multiplet are
\begin{align}
	\bar{L}^{\bold a}_j \Nabla_\a^j L_{\bold a}^i + \frac{8}{9} \Nabla_\a^i \log (|\cW|^2) = 0 ~, \qquad |\cW|^2 = {\cW^i \bar{\cW}_i}~.
\end{align}

Next, in complete analogy with the $\cN=2$ story described above, we utilise our local superconformal and $\sSU(3)_R \times \sSO(2)$ freedom to enforce the gauge 
\begin{align}
	|\cW| = 1 ~, \qquad L_{\bold a}^i = \d_{\bold a}^i ~,
\end{align}
which requires us to degauge the background from conformal to Einstein superspace, see section \ref{Section4} for details. One may then perform a similar analysis of the on-shell equations for the compensators given above, which we expect to indicate that the background is characterised by a cosmological constant. The specific details of this calculation will be provided elsewhere.

\noindent
{\bf Acknowledgements:}\\
We thank Gabriele Tartaglino-Mazzucchelli for email correspondence.
The work of SMK is supported in part by the Australian Research Council, project No. DP230101629.
The work of ESNR is supported by the Brian Dunlop Physics Fellowship.

\appendix

\section{The $\cN$-extended superconformal algebra} \label{AppendixA}

The conformal superspace approaches described in sections \ref{Section2.1} and \ref{Section3.1} are gauge theories of the superconformal algebra. In this appendix, we spell out our conventions for the $\cN$-extended superconformal algebra of Minkowski superspace \cite{HLS}, $\mathfrak{su}(2,2|\cN)$. 
Our normalisation of the generators of $\mathfrak{su}(2,2|\cN)$ is similar to \cite{FT}.
For comprehensive discussions of the superconformal transformations in superspace, see, e.g., 
\cite{Sohnius,Ferber,Park,KTh,Thesis}.

The conformal algebra, $\mathfrak{su}(2,2)$, consists of the translation $(P_a)$, Lorentz $(M_{ab})$, special conformal $(K_a)$ and dilatation $(\mathbb{D})$ generators. Amongst themselves, they obey the algebra
\begin{subequations} 
	\label{2.17}
	\begin{align}
		&[M_{ab},M_{cd}]=2\eta_{c[a}M_{b]d}-2\eta_{d[a}M_{b]c}~, \phantom{inserting blank space inserting} \\
		&[M_{ab},P_c]=2\eta_{c[a}P_{b]}~, \qquad \qquad \qquad \qquad ~ [\mathbb{D},P_a]=P_a~,\\
		&[M_{ab},K_c]=2\eta_{c[a}K_{b]}~, \qquad \qquad \qquad \qquad [\mathbb{D},K_a]=-K_a~,\\
		&[K_a,P_b]=2\eta_{ab}\mathbb{D}+2M_{ab}~.
	\end{align}
\end{subequations}

The $R$-symmetry group $\sU(\cN)_R$ is generated by the $\sU(1)_R$ $(\mathbb{Y})$ and $\sSU(\cN)_R$ $(\mathbb{J}^i{}_j)$ generators, which commute with all elements of the conformal algebra. Amongst themselves, they obey the commutation relations
\begin{align}
	[\mathbb{J}^{i}{}_j,\mathbb{J}^{k}{}_l] = \d^i_l \mathbb{J}^k{}_j - \d^k_j \mathbb{J}^i{}_l ~.
\end{align}

The superconformal algebra is then obtained by extending the translation generator to $P_A=(P_a,Q_\a^i,\bar{Q}^\ad_i)$ and the special conformal generator to $K^A=(K^a,S^\a_i,\bar{S}_\ad^i)$. The commutation relations involving the $Q$-supersymmetry generators with the bosonic ones are:
\begin{subequations} 
	\bea
	\big[M_{ab}, Q_\g^i \big] &=& (\s_{ab})_\g{}^\d Q_\d^i ~,\quad 
	\big[M_{ab}, \bar Q^\gd_i \big] = (\tilde{\s}_{ab})^\gd{}_\dd \bar Q^\dd_i~,\\
	\big[\mathbb{D}, Q_\a^i \big] &=& \hf Q_\a^i ~, \quad
	\big[\mathbb{D}, \bar Q^\ad_i \big] = \hf \bar Q^\ad_i ~, \\
	\big[\mathbb{Y}, Q_\a^i \big] &=&  Q_\a^i ~, \quad
	\big[\mathbb{Y}, \bar Q^\ad_i \big] = - \bar Q^\ad_i ~, \label{2.19c} \\
	\big[\mathbb{J}^i{}_j, Q_\a^k \big] &=&  - \d^k_j Q_\a^i + \frac{1}{\mathcal{N}} \d^i_j Q_\a^k ~, \quad
	\big[\mathbb{J}^i{}_j, \bar Q^\ad_k \big] = \d^i_k \bar Q^\ad_j - \frac{1}{\mathcal N} \d^i_j \bar Q^\ad_k ~,  \\
	\big[K^a, Q_\b^i \big] &=& -\ri (\s^a)_\b{}^\bd \bar{S}_\bd^i ~, \quad 
	\big[K^a, \bar{Q}^\bd_i \big] = 
	-\ri ({\s}^a)^\bd{}_\b S^\b_i ~.
	\eea
\end{subequations}
At the same time, the commutation relations involving the $S$-supersymmetry generators 
with the bosonic operators are: 
\begin{subequations}
	\bea
	\big [M_{ab} , S^\g_i \big] &=& - (\s_{ab})_\b{}^\g S^\b_i ~, \quad
	\big[M_{ab} , \bar S_\gd^i \big] = - (\ts_{ab})^\bd{}_\gd \bar S_\bd^i~, \\
	\big[\mathbb{D}, S^\a_i \big] &=& -\hf S^\a_i ~, \quad
	\big[\mathbb{D}, \bar S_\ad^i \big] = -\hf \bar S_\ad^i ~, \\
	\big[\mathbb{Y}, S^\a_i \big] &=&  -S^\a_i ~, \quad
	\big[\mathbb{Y}, \bar S_\ad^i \big] =  \bar S_\ad^i ~,  \label{2.20c}\\
	\big[\mathbb{J}^i{}_j, S^\a_k \big] &=&  \d^i_k S^\a_j - \frac{1}{\mathcal{N}} \d^i_j S^\a_k ~, \quad
	\big[\mathbb{J}^i{}_j, \bar S_\ad^k \big] = - \d_j^k \bar S_\ad^i + \frac{1}{\mathcal N} \d^i_j \bar S_\ad^k ~,  \\
	\big[ S^\a_i , P_b \big] &=& \ri (\s_b)^\a{}_\bd \bar{Q}^\bd_i ~, \quad 
	\big[\bar{S}_\ad^i , P_b \big] = 
	\ri ({\s}_b)_\ad{}^\b Q_\b^i ~.
	\eea
\end{subequations}
Finally, the anti-commutation relations of the fermionic generators are: 
\begin{subequations}
	\bea
	\{Q_\a^i , \bar{Q}^\ad_j \} &=& - 2 \ri \d^i_j (\s^b)_\a{}^\ad P_b=- 2 \ri \d^i_j  P_\a{}^\ad~, \\
	\{ S^\a_i , \bar{S}_\ad^j \} &=& 2 \ri  \d_i^j (\s^b)^\a{}_\ad K_b=2 \ri \d_i^j  K^\a{}_\ad
	~, \\
	\{ S^\a_i , Q_\b^j \} &=& \d_i^j \d^\a_\b \Big(2 \mathbb{D} + \frac{\mathcal{N}-4}{\mathcal{N}}\mathbb{Y} \Big) - 4 \d_i^j  M^\a{}_\b 
	+ 4 \d^\a_\b  \mathbb{J}^j{}_i ~, \\
	\{ \bar{S}_\ad^i , \bar{Q}^\bd_j \} &=& \d_j^i \d^\bd_\ad \Big(2 \mathbb{D} - \frac{\mathcal{N}-4}{\mathcal{N}}\mathbb{Y} \Big) + 4 \d_j^i  \bar{M}_\ad{}^\bd 
	- 4 \d_\ad^\bd  \mathbb{J}^i{}_j  ~. \label{2.21d}
	\eea
\end{subequations}
We emphasise that all (anti-)commutators not listed above vanish identically.

\section{Supergravity with cosmological term}
\label{AppendixB}

This appendix is devoted to a superspace description of $\cN=2$ supergravity with a cosmological term. We begin by coupling the background $\sSU(2)$ superspace reviewed in section \ref{section2.3} to an $\sSO(2)$ vector multiplet. We do this by introducing new covariant derivatives $\bm{\rm D}_A$
\begin{align}
	\bm{\rm D}_A = \mathscr{D}_A + \ri V_A \mathbb{Z}~,
\end{align}
where the $\sSO(2)$ generator $\mathbb{Z}$ acts on `second isospinors' $\psi_{\bold{a}}$ in accordance with the rule
\begin{align}
	\mathbb{Z} \psi_{\bold a} = \frac{\ri}{2} (\s^2)_{\bold a}{}^{\bold b} \psi_{\bold b}~, 
	\qquad \s^2 =\left(\begin{array}{c c}
		0 & - \ri \\
		\ri & 0 \\
	\end{array}\right) ~.
\end{align}
The resulting algebra of covariant derivatives then takes the form
\begin{subequations}
	\bea
	\{ \bm{\rm D}_\a^i , \bm{\rm D}_\b^j \}
	&=&
	4 S^{ij}  M_{\a\b} 
	+2\ve_{\a\b}\ve^{ij}Y^{\g\d}  M_{\g\d}  
	+2\ve^{ij} \ve_{\a\b}  \bar{W}_{\gd\dd} \bar{M}^{\gd\dd} 
	\non\\
	&&
	+2\ve_{\a\b}\ve^{ij}S^{kl} \mathbb{J}_{kl}
	+ 4Y_{\a\b}  \mathbb{J}^{ij} - 2\ri \ve_{\a \b} \ve^{ij} \mathcal{W} \mathbb{Z}
	~,
	\\
	\{ \bm{\rm D}_\a^i , \bar{\bm{\rm D}}^\bd_j \}
	&=&
	-2\ri\d^i_j\bm{\rm D}_\a{}^{\bd}
	+4\d^{i}_{j}G^{\d\bd}M_{\a\d}
	+4\d^{i}_{j}G_{\a\gd}\bar{M}^{\gd\bd}
	+8 G_\a{}^\bd J^{i}{}_{j}
	~,
	\eea
\end{subequations}
where the torsions $S^{ij}$, $Y_{\a \b}$, $G_{\aa}$ and $W_{\a \b}$ satisfy the usual Bianchi identities of $\sSU(2)$ superspace with the replacement $\mathscr{D}_A \rightarrow \bm{\rm D}_A$, while the field strength $\cW$ is a reduced chiral superfield
\begin{align}
	\bar{\bm{\rm D}}_i^\ad \cW = 0 ~, \qquad (\bm{\rm D}^{ij} + 4 S^{ij}) \cW = (\bar{\bm{\rm D}}^{ij} + 4 \bar{S}^{ij}) \bar{\cW}~.
\end{align}

To describe Poincar\'e supergravity, it is necessary to couple the background to two compensators, which are necessary to break the super-Weyl and $\sSU(2)_R$ symmetries. Below, the role of the former will be played by $\cW$, while we use a nonlinear multiplet $L_{\bold a}^i$ for the latter. In this supergeometry $L_{\bold a}^i$ is characterised by the nonlinear constraints
\begin{align}
	\text{det}(L) = 1 ~, \qquad L^{\bold a (i} \bm{\rm D}^{j}_\a L^{k)}_{\bold a} = 0~.
\end{align}
One may now utilise the local $\sSU(2)_R \times \sSO(2)$ freedom to fix the gauge $L_{\bold a}^i = \d_{\bold{a}}^i$, which leads to the consistency condition
\begin{align}
	\label{4.6}
	\F_\a^{(i,jk)} = - \frac{1}{2} \d^{(ij} V_{\a}^{k)}~.
\end{align}
Since this breaks the $\sSU(2)_R \times \sSO(2)$ symmetry down to its diagonal $\sSO(2)$ subgroup, the $\sSU(2)_R$ connection becomes auxiliary and should be separated. As a result, the spinor covariant derivatives degauge to
\begin{align}
	\label{4.7}
	\bm{\rm D}_\a^i = \bm{\cD}_\a^i - \chi_{\a j} \mathbb{J}^{ij} - \frac{1}{2} \d^{(ij} V_{\a}^{k)} \mathbb{J}_{jk} ~.
\end{align}
For simplicity, below we consider only geometries with $\chi_\a^i = 0$.\footnote{This turns out to be the equation of motion for $L_{\bold a}^i$ in Poincar\'e supergravity, see eq. \eqref{4.8b}.} 

Making use of eq. \eqref{4.7} allows one to compute the resulting algebra of spinor covariant derivatives. One finds
\begin{subequations}
	\begin{align}
		\{ \bm{\cD}_\a^i , \bm{\cD}_\b^j \} &= - \d^{(ij} V_{(\a}^{k)} \bm{\cD}_{\b) k} + 4 S^{ij} M_{\a \b} + 2 \ve_{\a \b} \ve^{i j} \Big ( Y^{\g \d} M_{\g \d} + \bar{W}^{\gd \dd} \bar{M}_{\gd \dd} - \ri \bar{\cW} \mathbb{Z} \Big ) ~, \\
		\{ \bm{\cD}_\a^i , \bar{\bm{\cD}}_{\bd}^j \} &= - \frac{1}{2} \d^{(ij} V_\a^{k)} \bar{\bm{\cD}}_{\bd k} + \frac{1}{2} \d^{(ij} \bar{V}_\bd^{k)} \bm{\cD}_{\a k} + 2 \ri \ve^{ij} \bm{\cD}_{\a \bd} - 4 \ve^{ij} \Big( G^{\b}{}_{\bd} M_{\a \b} - G_{\a}{}^{\ad} \bar{M}_{\ad \bd} \Big)~,
	\end{align}
\end{subequations}
where $\bm{\cD}_\aa = \bm{\rm D}_\aa$, and the dimension-1 torsions $S^{ij}$, $Y_{\a \b}$, $G_\aa$ and $\cW$ are expressed in terms of $V_\a^i$ as follows
\begin{subequations}
	\label{4.9}
	\begin{align}
		S^{ij} &= -\frac{1}{24} \d^{ij} \bm{\cD}^\a_k V_\a^k - \frac{1}{12} \d^{k(i} \bm{\cD}_k^\a V_\a^{j)} - \frac{1}{24} V^{\a i} V_{\a}^j - \frac{1}{72} \d^{ik} \d^{jl} V^{\a}_k V_{\a l} ~, \\
		Y_{\a \b} &= \frac{1}{36} \Big( \d_{ij} \bm{\cD}_{(\a}^i V_{\b)}^j - V_{(\a}^i V_{\b) i} \Big)~, \\
		G_{\aa} &= \frac{1}{144} \Big( \d^{ij} (\bm{\cD}_{\a i} \bar{V}_{\ad j} + \bar{\bm{\cD}}_{\ad i} V_{\a j} ) - \frac{14}{3} V_\a^i \bar{V}_{\ad i}\Big)~, \\
		\cW &= - \frac{1}{4} \bm{\cD}^\a_i V_\a^i ~.
	\end{align}
\end{subequations}
This means that the resulting geometry is described solely in terms of $V_\a^i$ and the super-Weyl tensor $W_{\a \b}$, which obey the Bianchi identities
\begin{subequations}
	\begin{align}
		&\bar{\bm{\cD}}_\ad^i W_{\a \b}  = 0 ~, \qquad \bm{\cD}^\b_i W_{\a \b} + 2 \bar{\bm{\cD}}^\ad_i G_{\aa} - \frac{1}{3} \bm{\cD}_{\a}^j \bar{S}_{ij} + \frac{1}{6} \d_{(ij} V_{\a k)} \bar{S}^{jk} = 0 ~, \\
		&\bar{\bm{\cD}}_i^\ad \cW = 0 ~, \qquad \Big(\bm{\cD}^{ij} + 4 S^{ij} \Big) \cW = \Big(\bar{\bm{\cD}}^{ij} + 4 \bar{S}^{ij} \Big) \bar{\cW}~, \\
		& \d^{k(i} \bm{\cD}_{(\a k} V_{\b)}^{j)} + \frac{1}{3} \d^{k(i} \bm{\cD}_{(\a}^{j)} V_{\b) k} = 0~, \\
		& \d_{ij} (\bm{\cD}_\a^k \bar{V}_{\ad i} + \bar{\bm{\cD}}_{\ad i} V_{\a}^i ) + 2 \d_{k (i} \bm{\cD}_\a^k \bar{V}_{\ad j)} + 2 \bar{\bm{\cD}}_{\ad (i} V_{\a j)} + 2 V_{\a(i} \bar{V}_{\ad j)} - \frac{2}{3} \d_{k(i} V_\a^k \bar{V}_{\ad j)}= 0~, \\
		& \d^{k(i} (\bm{\cD}_\a^{j)} \bar{V}_{\ad k} + \bar{\bm{\cD}}_\ad^{j)} V_{\a k}) - \frac{4}{3} \d^{k(i} \d^{j)l} V_{\a k} \bar{V}_{\ad l} = 0~.
	\end{align}
\end{subequations}

Next, by performing a straightforward application of the Jacobi identity for $\bm{\cD}_A$, one may extract the scalar curvature for this supergeometry. It is given by
\begin{align}
	\cR &= - \frac{1}{4} \Big( \bm{\cD}_\aa + \frac{1}{108} V_\a^i \bar{V}_{\ad i} - 2 G_{\aa} \Big) G^{\aa} - \frac{1}{288} \Big [ 
	\Big( \bm{\cD}^{ij} + 2 V^{\a i} V_{\a}^j + 12 \d^{(ij} \bm{\cD}^\a_k V_{\a}^{k)} \non \\
	&\phantom{=} \qquad - \d^{ik} \d^{jl} V^\a_k V_{\a l} + 3 \bar{\cW} \d^{ij} \Big) \bar{S}_{ij} + 3 \d^{(ij} \bm{\cD}^\a_i \bar{S}_{jk} V_\a^{k)} + \text{c.c.}
	\Big ] ~.
\end{align}
We see that it contains a $\cW$-dependent sector, which can be further simplified via eq. \eqref{4.9}
\begin{align}
	-\frac{1}{96} \Big(\cW \d^{ij} S_{ij} + \bar{\cW} \d_{ij} \bar{S}^{ij} \Big) = - \frac{1}{72} \cW \bar{\cW} + \dots~,
\end{align} 
where the ellipses denote terms irrelevant to our analysis. Finally, by making use of the super-Weyl symmetry, one fixes the gauge
\begin{align}
	\cW = \text{const}
\end{align}
leading to a negative cosmological constant. Our analysis is in agreement with, and may be understood as a superspace analogue of, the results obtained in \cite{deWvanHvanP}.

\begin{footnotesize}

\end{footnotesize}

\end{document}